\documentclass[cameraready]{Interspeech}
\usepackage{textcomp}
\usepackage{comment}
\usepackage{subcaption}
\usepackage{tabularx}
\usepackage{amsmath}
\usepackage{amssymb}
\usepackage{placeins}
\usepackage{float}
\usepackage{needspace}
\usepackage{hyperref}
\usepackage{multicol}
\usepackage{multirow}
\usepackage{cite}
\usepackage{graphicx}
\usepackage{url}
\usepackage{comment}
\usepackage[hang,flushmargin]{footmisc}
\graphicspath{{figs/}{figures/}{figures/results/}}

\title{Dimensionality-Aware Anomaly Detection in Learned Representations of Self-Supervised Speech Models}

\author[affiliation={1,3},
correspondingauthor,
orcid=0009-0008-0680-1817, 
]{Sandra}{Arcos-Holzinger}

\author[affiliation={1}, 
orcid=0000-0003-0885-0643, 
]{Sarah M.}{Erfani}

\author[affiliation={2},
orcid=0000-0002-3769-3811
]{James}{Bailey}

\author[affiliation={3}, 
orcid=0000-0001-5976-0897, 
]{Sanjeev}{Khudanpur}

\email{sarcosh1@jhu.edu
}

\address{
    $^1$ University of Melbourne, School of Computing and Information Systems, Australia\\
    $^2$ Monash University, Department of Data Science and Artificial Intelligence, Australia\\
    $^3$ Johns Hopkins University, Center for Language and Speech Processing, USA 
}


\begin{document}

\maketitle

\begin{abstract}
Self-supervised speech models (S3Ms) achieve strong downstream performance, yet their learned representations remain poorly understood under natural and adversarial perturbations. Prior studies rely on representation similarity or global dimensionality, offering limited visibility into local geometric changes. We ask: how do perturbations deform local geometry, and do these shifts track downstream automatic speech recognition (ASR) degradation? To address this, we present \textit{GRIDS}, a framework using Local Intrinsic Dimensionality (LID) across layer-wise representations in WavLM and wav2vec~2.0.  We find that LID increases for all low signal-to-noise ratio (SNR) perturbations and diverges at high SNR: benign noise converges toward the clean profile, while adversarial inputs retain early-layer LID elevation. We show LID elevation co-occurs with increased WER, and that layer-wise LID features enable anomaly detection (AUROC~0.78–1.00), opening the door to transcript-free monitoring in S3Ms.
\end{abstract}

\section{Introduction}\label{sec:intro}
Self-supervised speech representations are an active area of research, with prior work motivating a deeper analysis of information encoding in model layers and understanding the behavior of representations under distributional shifts \cite{Mohamed2022-ssl-review}. Existing work has sought to enhance the noise robustness of self-supervised speech models (S3Ms), as well as characterize their adversarial vulnerability \cite{hsu2021robust_wav2vec2,huang2022distortion_domain_adapt,wang2022speech_reconstruction,zhu2022noise_robust_ssl,wu2021adv_vulnerability_speech}. Analysis of learned representations indicates that information content varies substantially between layers in S3Ms, and that extracting features from the last layers may not be optimal for tasks that require phonetic or word-related information \cite{Mohamed2022-ssl-review}. Motivated by these observations, we use Local Intrinsic Dimensionality (LID) as a layer-wise geometric diagnostic to quantify how benign and adversarial distortions deform local neighborhood structure.

LID characterizes the geometric properties of learned representations by quantifying how rapidly the number of neighboring samples grows as the radius around a reference sample increases~-- a quantity known as the local rate of probability mass expansion~\cite{lid-amsaleg1028,lid-bailey2021}. It has been used in a range of contexts, including outlier detection \cite{Houle2018} and neural networks~\cite{lid-bailey2019,lid-Ansuini2019}, and has proven effective in revealing how adversarial perturbations alter learned representations in the image \cite{huang2025detectingbackdoorsamplescontrastive-localsubspace-LID-JamesB, lid-Pope2021,lid-Ma2018,lid-Gong2019,lid-weerasinghe2022} and text \cite{ruppik2025less} domains, where perturbed samples consistently exhibit higher LID values corresponding to low-density regions in the pixel or deep feature space. In spite of this, and to the best of our knowledge, LID has not been studied in the context of speech representations in S3Ms. Intuitively, LID measures the relative rate at which the neighborhood around a nearby sample grows: expanding rapidly in high-dimensional regions, and slowly in low-dimensional ones. 

We use \emph{manifold} to denote the low-dimensional structure that clean-speech representations occupy within the high-dimensional embedding space~-- the \emph{ambient dimension}~-- of a model's hidden layers, consistent with the manifold hypothesis that high-dimensional data concentrates on or near low-dimensional manifolds~\cite{NIPS2010_8a1e808b, fefferman2016testing}. The true \emph{degrees of freedom} of the data, i.e., the number of independent directions along which representations locally vary, are typically much fewer than the ambient dimension. We use \emph{geometry} to denote the measurable local neighborhood properties of that space~-- such as distances to nearest neighbors, local density, expansion rate, and effective local dimensionality~-- that characterize how representations are arranged and how perturbations deform those arrangements. 

LID is a geometric statistic in this sense: it estimates the effective local dimensionality around a sample, which we leverage to quantify shifts between natural and adversarial conditions relative to the clean manifold structure. 
A progressive decrease in LID across layers reflects representational compression~-- a reduction in effective degrees of freedom as the model abstracts from acoustic to linguistic structure. This abstraction hierarchy has been independently established through a layer-wise representation analysis~\cite{Pasad_2021-cca-layerwise-Livescu}, and our LID trajectories in S3Ms provide convergent geometric evidence for the same progression.
If geometric structure is integral to how S3Ms encode information across layers, then geometric changes due to perturbations should impact downstream performance degradation. This hypothesis motivates understanding changes in LID to task-level metrics such as word error rate (WER) in Automatic Speech Recognition (ASR).

We extend this geometric approach to ASR-based S3Ms to investigate how layer-wise representations change under adversarial and benign acoustic perturbations. This extension presents a non-trivial challenge: unlike prior LID applications ~\cite{ma2018characterizing} where each input yields a single feature vector, S3Ms generate variable-length frame-level embeddings per utterance. This requires careful aggregation of the embeddings to ensure a stable and reliable LID estimation. Furthermore, S3M features are learned via self-supervision and remain task-agnostic~\cite{Mohamed2022-ssl-review}. Whether LID is equally effective in S3Ms as shown in supervised settings~\cite{ma2018characterizing} remains an open empirical question. Consequently, we utilize LID as a diagnostic of S3M transformer layers, empirically linking intrinsic geometry in the learned representations of S3Ms to downstream robustness. 

We refer to our framework as \textit{Geometric Robustness via Intrinsic Dimensionality in Speech (GRIDS)}. We apply it to WavLM~\cite{Chen2021WavLMLS} and wav2vec~2.0~\cite{baevski2020wav2vec}, and make the following contributions:

\begin{enumerate}[label=(\roman*)]
    \item \textit{LID--S3M geometric analysis:} measurement of local geometric changes in learned representations across transformer layers for clean, benign, and adversarial conditions;
    \item \textit{LID--ASR monitoring:} empirical evidence showing that increases in LID positively associate with WER, linking geometric changes to downstream ASR degradation;
    \item \textit{LID--AD for anomaly detection:} 12 LID-derived features from (i) for adversarial vs. benign classification.
\end{enumerate}

\section{Related Work}\label{sec:related}
\subsection{Representation Analysis and Dimensionality in Self-Supervised Models}
Recent work on transformer representation geometry has shown that learned features organize on low-dimensional curved manifolds within a model's hidden layers, and that attention heads manipulate this manifold structure as a computational mechanism~\cite{gurnee2026modelsmanipulatemanifoldsgeometry_anthropic}. This provides empirical support for the manifold hypothesis and reinforces the value of geometric analysis -- such as intrinsic dimensionality estimation -- for understanding how representations encode structure across layers. 

Layer-wise representation analysis has proven to be instrumental in the interpretation of self-supervised speech models, providing information on how different layers encode phonetic, syntactic, and semantic information~\cite{pasad2023comparative}. Prior work has primarily employed similarity-based metrics to study these relationships, each providing a complementary lens on how representations evolve across a network. Canonical Correlation Analysis (CCA) and Centred Kernel Alignment (CKA), for instance, measure \emph{how similar} representations are between layers. CCA and its variants identify linear relationships between representations by finding optimal alignment transformations~\cite{raghu2017svcca-brno-CKA-CCA_CNN-LSTM-transformer,morcos2018insights} (with its non-linear extensions~\cite{cca-nonlinear1, cca-nonlinear2, cca-nonlinear3, cca-nonlinear4, cca-nonlinear5, cca-nonlinear6}), whereas CKA extends this to high-dimensional spaces by directly comparing similarity structures rather than feature values~\cite{kornblith2019similarity-cka-Hinton}. For speech models specifically, ~\cite{Pasad_2021-cca-layerwise-Livescu} applied these metrics to S3Ms such as wav2vec~2.0~\cite{baevski2020wav2vec} and HuBERT~\cite{Hsu_2021-HuBERT}, revealing a hierarchical flow of information from acoustic to linguistic abstractions across layers.

While alignment-based similarity measures provide valuable insight into \emph{representation flow}, they do not directly quantify \emph{representation geometry} or detect when individual samples deviate from the learned manifold. Rank-based approaches~\cite{rankme-garrido-lecun-2023} operating on the singular value spectrum of embedding matrices address dimensionality more directly. In ~\cite{aldeneh2024rankme}, it was shown that the effective rank of layer-wise embeddings in HuBERT correlates with downstream task performance, demonstrating that global measures of dimensionality are informative for assessing S3M representation quality. However, rank captures dataset-level dimensional collapse -- the degree to which representations fail to occupy the full embedding space -- and cannot predict which layer performs best on a given downstream task~\cite{aldeneh2024rankme}. This suggests that global dimensionality measures alone may not fully capture local variation in learned representations across S3M layers. To date, dimensionality-based approaches have not been extended to robustness analysis under perturbation in S3Ms, nor used to relate geometric properties of learned representations to downstream task degradation.

\subsection{Local Intrinsic Dimensionality as a 
Geometric Diagnostic}
LID offers a complementary perspective: rather than measuring a global rank or how similar representations are between layers, LID estimates the dimensional properties of representation manifolds in the neighborhood of individual samples, capturing \emph{how complex} the geometry of the representation manifold is within each layer. LID thus quantifies the effective number of degrees of freedom -- or intrinsic dimension -- required to describe the local geometry of representations within each layer. It operates directly on each layer's representation space, estimating local geometric complexity through nearest-neighbor distance and expansion rate -- making it uniquely suitable for identifying outliers and characterizing how adversarial or perturbed inputs distort the representation manifold. The utility of LID to detect adversarial examples in the latent space of vision models was first demonstrated in~\cite{ma2018characterizing}. It showed that adversarial samples exhibit abnormally high LID values, which correspond to low-probability and high-dimensional regions of feature space that require more dimensions to describe their local neighborhood structure.

Subsequent studies have also extended LID to other domains: \cite{huang2025detectingbackdoorsamplescontrastive-localsubspace-LID-JamesB} applied it for backdoor image detection, showing that poisoned samples produce distinctive subspace anomalies, while \cite{ruppik2025less} analyzed contextual language models using LID to characterize perturbation effects in text representations. Although LID has been explored in vision, text, and even conventional audio analysis for voice detection \cite{Liu2018Voice}, and classification of audio with adversarial examples in neural networks (i.e., AlexNet, GoogLeNet) and a linear SVM \cite{Esmaeilpour2019-adversarial-audio-dnns}, it has not been applied in the analysis of S3Ms such as WavLM~\cite{Chen2021WavLMLS} and wav2vec~2.0~\cite{baevski2020wav2vec}. 

To the best of our knowledge, no prior work has performed a layer-wise geometric analysis using LID across transformer representations in S3Ms or investigated its potential as a dimensionality-aware diagnostic for robustness monitoring. We address this gap by conducting a layer-wise LID analysis of WavLM and wav2vec~2.0 under matched target-SNR adversarial and benign perturbations, examining how intrinsic dimensionality changes relate to robustness and perturbation sensitivity in S3Ms. Our work adds a complementary local and geometric lens to the analysis of S3M representations: where existing approaches measure representational similarity or global dimensionality, LID captures local distortion -- enabling per-sample, per-layer detection of geometric anomalies that global measures cannot resolve.

\section{Methodology}\label{sec:method}
\begin{figure*}[t!]
  \centering
  \includegraphics[width=\textwidth]{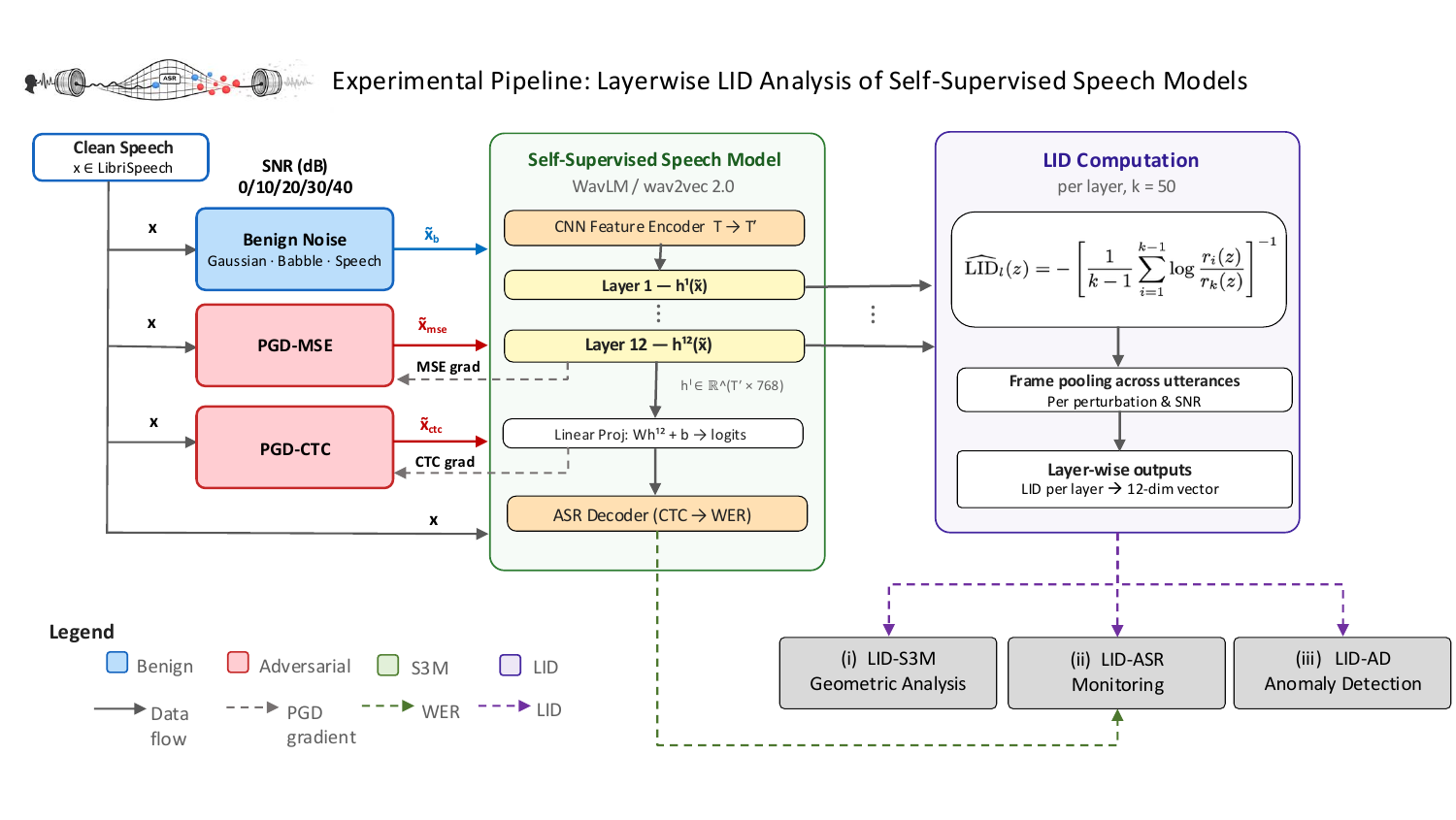}
  \caption{Experimental end-to-end pipeline and overview of our \textit{GRIDS} framework. Clean and perturbed utterances, including benign noise (Gaussian, babble, speech) and PGD-adversarial attacks (MSE, CTC), are independently passed through WavLM and wav2vec~2.0 under matched target-SNR conditions. Layer-wise LID estimates support three analyses: (i)~LID--S3M geometric analysis, (ii)~LID--ASR monitoring, and (iii)~LID--AD anomaly detection.}
  \label{fig:pipeline}
\end{figure*}

\subsection{LID Layer-wise Analysis in WavLM and wav2vec~2.0}
Our goal is to quantify how benign and adversarial perturbations deform the local geometry of S3M representations across transformer layers, and to test whether these geometric shifts track downstream ASR degradation and support anomaly detection. Figure~\ref{fig:pipeline} summarizes our \textit{GRIDS} framework. Clean LibriSpeech utterances and their perturbed counterparts, including benign noise (blue blocks) and PGD adversarial attacks (red blocks), are passed independently through WavLM and wav2vec~2.0 under matched target-SNR conditions (green block). For the LID--S3M and LID--ASR analyses, we use a kNN pipeline: local LID is estimated per frame, reduced to a layer-level scalar by a harmonic mean over valid local estimates, and then reduced across layers by a second harmonic mean when an overall condition-level summary is needed.

Let $x\in\mathbb{R}^T$ be a waveform and let $h^l(x)\in\mathbb{R}^{T'\times d}$ denote its hidden representation at transformer layer $l$, where $T'$ is the number of frames after convolutional subsampling and $d$ is the hidden size. We index layers by $l\in\{1,\dots,L\}$, with $L=12$ for the base WavLM and wav2vec~2.0 encoders considered here. Following a Levina--Bickel style MLE estimator~\cite{NIPS2004_74934548}, we compute \emph{local} LID for a frame-level embedding $z=h_t^l(x)\in\mathbb{R}^d$ from its first $k-1$ nearest neighbors relative to the $k$-th-neighbor radius:

\begin{equation}
\color{black}
\widehat{\mathrm{LID}}_l(z) = -\left[\frac{1}{k-1}
\sum_{i=1}^{k-1} \log \frac{r_i(z)}{r_k(z)}
\right]^{-1}.
\end{equation}

\noindent where $r_i(z)$ is the Euclidean distance from $z$ to its $i$-th nearest neighbor and $k$ is the neighborhood size. Distances are computed within each layer and perturbation condition, with embeddings standardized to zero mean and unit variance separately for each layer-condition pair before kNN search. In the implementation, numerically unstable local estimates are clamped to a finite range for stability, so $\Delta\mathrm{LID}$ should be interpreted as a change in normalized local neighborhood geometry rather than as a comparison of raw feature norms across conditions. We define a condition $c = (\text{model}, \text{perturbation type}, \text{SNR}$) as a tuple specifying the experimental setting under which embeddings are extracted. For each condition $c$, we pool frame embeddings across utterances within that condition and aggregate valid local LID estimates with a harmonic mean:
\begin{equation}
\color{black}
\mathrm{LID}_{l}^{(c)}=
\left[
\frac{1}{|S_l^{(c)}|}
\sum_{z\in S_l^{(c)}}\frac{1}{\widehat{\mathrm{LID}}_l(z)}
\right]^{-1},
\label{eq:global_lid_hmean}
\end{equation}
where $S_l^{(c)}$ denotes the pooled set of layer-$l$ frame embeddings for condition $c$. This pooled estimator is used for the condition-level LID--S3M and LID--ASR analyses.

For per-layer perturbation effects, we report
\begin{equation}
\color{black}
\Delta\mathrm{LID}_l
=
\mathrm{LID}_{l}^{(\mathrm{pert})}
-
\mathrm{LID}_{l}^{(\mathrm{clean})}.
\label{eq:delta_lid_def}
\end{equation}

For LID--ASR analysis, we further define the cross-layer summary for condition $c$ as
\begin{equation}
\color{black}
\mathrm{LID}_{\mathrm{overall}}^{(c)}
=
\left[
\frac{1}{L}
\sum_{l=1}^{L}
\frac{1}{\mathrm{LID}_{l}^{(c)}}
\right]^{-1}.
\label{eq:overall_lid_def}
\end{equation}

It follows that the perturbation-induced LID shift can be captured by
\begin{equation}
\color{black}
\Delta\mathrm{LID}
=
\mathrm{LID}_{\mathrm{overall}}^{(\mathrm{pert})}
-
\mathrm{LID}_{\mathrm{overall}}^{(\mathrm{clean})}.
\label{eq:delta_lid_overall_def}
\end{equation}

In the current pipeline, $k$ is not treated as a purely fixed hyperparameter. For each condition, we evaluate a grid of neighborhood sizes and compute $\Delta\mathrm{LID}(k)$ from Eq.~\eqref{eq:delta_lid_overall_def}. We retain the $k$ values whose overall $\Delta\mathrm{LID}$ lies within a fixed fraction of the best value, and choose the candidate with minimum across-layer standard deviation of $\Delta\mathrm{LID}$ (stability across layers). When multiple candidates achieve this minimum, the larger $\Delta\mathrm{LID}$ is chosen (discriminability against the clean baseline); when both criteria are matched, a smaller $k$ is chosen, given that a larger neighborhood would not improve stability or discriminability and would only blur the local estimate.

For anomaly detection, we switch from pooled condition-level summaries to utterance-level summaries. Each utterance $x_i$ is represented by the 12-dimensional LID feature vector
\begin{equation}
\color{black}
\mathbf{v}_i
=
\bigl[\mathrm{LID}_{i,1},\ldots,\mathrm{LID}_{i,12}\bigr]^\top
\in\mathbb{R}^{12},
\label{eq:lid_feature_vector}
\end{equation}
where $\mathrm{LID}_{i,l}$ is the utterance-level harmonic-mean LID computed from the frame embeddings of utterance $i$ at layer $l$. Accordingly, the geometry and WER analyses use pooled condition-level LID summaries, whereas anomaly detection uses utterance-level LID vectors.

\subsection{Perturbation Generation}
This section provides details on the generation of perturbations used in our experiments. 

\subsubsection{Target-SNR constraints and shared scaling}
We compare benign and adversarial perturbations under controlled target-SNR conditions by enforcing a per-utterance SNR constraint on the perturbation energy. Given a clean waveform $x$ and additive perturbation $\delta$, we define
\begin{equation}
\mathrm{SNR}_{\mathrm{dB}}(x,\delta)=20\log_{10}\frac{\lVert x\rVert_2}{\lVert \delta\rVert_2}.
\label{eq:snr_def}
\end{equation}
For a target SNR $s$ (in dB), we constrain $\lVert \delta \rVert_2\le\epsilon_{\mathrm{SNR}}(x,s)$ with
\begin{equation}
\epsilon_{\mathrm{SNR}}(x,s)=\lVert x\rVert_2\,10^{-s/20}.
\label{eq:eps_snr}
\end{equation}

For benign perturbations, we first sample a raw waveform $\delta_{\mathrm{raw}}$ and then rescale it to the target SNR as
\begin{equation}
\delta=\epsilon_{\mathrm{SNR}}(x,s)\,\frac{\delta_{\mathrm{raw}}}{\lVert\delta_{\mathrm{raw}}\rVert_2}.
\label{eq:snr_rescale}
\end{equation}

For Gaussian noise, the target SNR is enforced as a constraint (cap) rather than an exact rescaling. The raw perturbation is projected into the $\ell_2$ ball:
\begin{equation}
\delta=\delta_{\mathrm{raw}} \cdot\min\left(1,\frac{\epsilon_{\mathrm{SNR}}(x,s)}{\lVert\delta_{\mathrm{raw}}\rVert_2}\right).
\label{eq:snr_project}
\end{equation}
All perturbed waveforms are finally clipped by $\tilde{x}=\mathrm{clip}(x+\delta,-1,1)$ to ensure a valid waveform range.

\subsubsection{Benign and Adversarial Perturbations}
Perturbations are generated from LibriSpeech~\cite{LibriSpeechDataset} clean utterances using our end-to-end \textit{GRIDS} pipeline shown in Figure~\ref{fig:pipeline}. The paired utterance subset and train/evaluation protocol are specified in Section~\ref{sec:experimental_config}. We consider three benign perturbations~--~Gaussian noise, Noizeus babble noise~\cite{NoizeusDataset}, and a secondary LibriSpeech competing-talker utterance, all rescaled to the target SNR (Eq.~\eqref{eq:snr_rescale}; Eq.~\eqref{eq:snr_project} for the Gaussian cap)~--~and PGD adversarial perturbations~\cite{madry2019-PGD} under an $\ell_2$-bounded budget calibrated to the same target SNR levels. Per-perturbation settings are summarized in Table~\ref{tab:perturb_params}. Adversarial examples use one of two objectives: Mean Squared Error (MSE-loss), Eq.~\eqref{eq:mse-loss} or Connectionist Temporal Classification (CTC-loss), Eq.~\eqref{eq:ctc-loss}~\cite{graves2006-ctc}. MSE-PGD serves as a baseline that maximally distorts final-layer hidden representations (non-task-specific), whereas CTC-PGD is task-aligned to ASR.

\begin{table}[ht]
\centering
\footnotesize
\setlength{\tabcolsep}{2pt}
\caption{Perturbation-generation settings used in our experiments with target SNR range $\{0,10,20,30,40\}$\,dB.}
\label{tab:perturb_params}
\begin{tabularx}{\columnwidth}{@{}lX@{}}
\toprule
\textbf{Perturbation }& \textbf{Settings}\\
\midrule
\textbf{(1) Gaussian noise} & i.i.d.\ $\mathcal{N}(0,\sigma^2)$, $\sigma{=}0.01$; SNR cap via $\ell_2$ projection in Eq.~\eqref{eq:snr_project}; clip $[-1,1]$ \\
\textbf{(2) Babble noise }& Noizeus babble-noise clips; full overlap; rescaled to the target SNR in Eq.~\eqref{eq:snr_rescale}\\
\textbf{(3) Speech noise} & LibriSpeech competing talker; full overlap; rescaled to the target SNR in Eq.~\eqref{eq:snr_rescale}\\
\textbf{(4) Adversarial PGD} & $\ell_2$ constraint Eq.~\eqref{eq:eps_snr}; 300 iterations; base step size $\eta{=}0.01$ scaled by $\epsilon_{\mathrm{SNR}}$; post-loop rescale to exhaust $\epsilon_{\mathrm{SNR}}$; clip $[-1,1]$ \\
\bottomrule
\end{tabularx}
\end{table}

\subsubsection{Adversarial examples and MSE loss}
We use MSE loss as the squared Frobenius norm of the difference between final-layer hidden representations:
\begin{equation}
\mathcal{L}_{\mathrm{MSE}}(x,\delta)
=
\frac{1}{N}
\left\lVert h^{L}(x+\delta)-h^{L}(x)\right\rVert_F^2,
\label{eq:mse-loss}
\end{equation}

\noindent where $h^{L}(x) \in \mathbb{R}^{T' \times d}$ denotes the hidden representations of the final transformer block (i.e., after the layer norm) and $\lVert \cdot \rVert_F$ denotes the Frobenius norm. $T'$ is the number of time frames after the convolutional subsampling stage in the model. The factor $N = T' \times d$ normalizes the loss by averaging over all temporal frames and feature dimensions.

\subsubsection{Adversarial examples and CTC loss}
\newcommand{\blank}{\varnothing}
\newcommand{\Vocab}{\mathcal{V}} 
We use CTC loss as an adversarial objective to maximize transcription disruption under the same SNR-constrained $\ell_2$ budget. Let $\Vocab$ denote the output token vocabulary and let $\Vocab'=\Vocab\cup\{\blank\}$ denote the CTC-extended vocabulary including the blank symbol. For an utterance $x$, the final-layer representation $h^L(x)\in\mathbb{R}^{T'\times d}$ induces frame-level logits
\begin{equation}
s_t(x)=W h_t^L(x)+b,\qquad t=1,\ldots,T',
\label{eq:ctc_logits}
\end{equation}
where $W\in\mathbb{R}^{|\Vocab'|\times d}$ and $b\in\mathbb{R}^{|\Vocab'|}$. The posterior probability of token $k\in\Vocab'$ at frame $t$ is
\[
p(k\mid x,t)=\mathrm{softmax}(s_t(x))_k.
\]

A CTC path is a frame-level label sequence $\pi=(\pi_1,\ldots,\pi_{T'})\in(\Vocab')^{T'}$ with path probability
\[
p(\pi\mid x)=\prod_{t=1}^{T'} p(\pi_t\mid x,t).
\]

Let $y=(y_1,\ldots,y_U)\in\Vocab^U$ denote the target transcription, with $U\le T'$. The collapse operator $\mathcal{B}:(\Vocab')^{T'}\rightarrow \Vocab^{\le T'}$ removes blank symbols and merges consecutive repeated labels. The conditional probability of $y$ is therefore
\begin{equation}
p(y\mid x)=\sum_{\pi\in\mathcal{B}^{-1}(y)} p(\pi\mid x),
\label{eq:ctc_conditional}
\end{equation}
and the CTC loss is
\begin{equation}
\mathcal{L}_{\mathrm{CTC}}(x,y)=-\log p(y\mid x).
\label{eq:ctc-loss}
\end{equation}

The untargeted adversarial objectives are therefore
\begin{align}
\max_{\lVert \delta \rVert_2 \le \epsilon_{\mathrm{SNR}}(x,s)}
\mathcal{L}_{\mathrm{MSE}}(x,\delta),
\label{eq:mse-untargeted}
\\
\max_{\lVert \delta \rVert_2 \le \epsilon_{\mathrm{SNR}}(x,s)}
\mathcal{L}_{\mathrm{CTC}}(x+\delta,y).
\label{eq:ctc-untargeted}
\end{align}

We optimize $\delta$ with $\ell_2$-PGD using a normalized-gradient update and projection to the SNR budget:
\begin{equation}
\delta_{t+1}=\Pi_{\epsilon_{\mathrm{SNR}}}\left(\delta_t+\alpha_t\frac{\nabla_\delta \mathcal{L}(x+\delta_t)}{\lVert\nabla_\delta \mathcal{L}(x+\delta_t)\rVert_2}\right),
\label{eq:pgd_update}
\end{equation}
where $\Pi_{\epsilon_{\mathrm{SNR}}}$ denotes projection onto the $\ell_2$ ball of radius $\epsilon_{\mathrm{SNR}}(x,s)$ defined in Eq.~\ref{eq:eps_snr}. We use a per-utterance step schedule $\alpha_t=\eta\,\epsilon_{\mathrm{SNR}}(x,s)\,(1+2e^{-t/20})$ with $\eta=0.01$. After the final iteration, we rescale $\delta$ to exhaust the budget (i.e., $\lVert \delta\rVert_2=\epsilon_{\mathrm{SNR}}$) and clip to $[-1,1]$.

\subsection{Classification with LID derived features}
For anomaly detection, we construct a 12-dimensional LID feature vector per utterance, as defined in Eq.~\eqref{eq:lid_feature_vector}, and train a lightweight classifier on these utterance-level summaries. This classifier stage is distinct from the pooled condition-level LID analysis used for geometry and WER, and its evaluation protocol is detailed in Section \ref{sec:experimental_config}. In this section, we define the attack-effectiveness metric used to contextualize classifier difficulty and then describe the LID-based anomaly-detection task.
\subsubsection{Attack success rate}\label{sec:attack_success_rate}
To quantify perturbation effectiveness, we define the attack success rate $\mathrm{SR}_{\gamma,\tau}$ as the empirical fraction of utterances for which the perturbed WER satisfies two joint conditions: (1) it reaches at least an absolute threshold $\tau$, and (2) the increase over the clean WER is at least $\gamma$:

\begin{equation}
\color{black}
\begin{aligned}
\mathrm{SR}_{\gamma,\tau}
&=
\frac{1}{n}\sum_{i=1}^{n}
\mathbf{1}\left[
\mathrm{WER}_{\mathrm{pert},i}\ge\tau
\right.
\\
&\qquad\left.
\land\
\bigl(\mathrm{WER}_{\mathrm{pert},i}-\mathrm{WER}_{\mathrm{clean},i}\bigr)\ge\gamma
\right],
\end{aligned}
\label{eq:success_rate_and_wer}
\end{equation}

where $n$ is the number of utterances and $\mathbf{1}[\cdot]$ is the indicator function. We write $\mathrm{WER}_{\mathrm{pert},i}$ to emphasize that the same definition is applied consistently across perturbation types, while anomaly-detection results in Section \ref{subsec:anomaly_detection} focus specifically on adversarial-versus-benign separation.

\subsubsection{LID-based anomaly detection}\label{sec:anomaly_detection_method}

For anomaly detection, each utterance $x_i$ is represented by the feature vector $\mathbf{v}_i$ defined in Eq.~\eqref{eq:lid_feature_vector}. For each model, attack objective, and SNR, we form a binary task with positives given by adversarial utterances and negatives given by the pooled benign noise set (Gaussian, babble, and speech) at the same SNR. We train a logistic-regression classifier and evaluate it with 5-fold grouped cross-validation keyed by normalized utterance ID, so that multiple variants of the same underlying utterance never appear in both train and test folds. We report AUROC, AUPRC, and FPR at TPR $=0.95$ from concatenated out-of-fold scores.

\section{Experimental Configuration}\label{sec:experimental_config}
We draw utterances from LibriSpeech \texttt{test-clean}~\cite{LibriSpeechDataset}, 
selecting 40 speakers at random retaining only utterances of 5-10\,s duration (16\,kHz sampling frequency). We follow this approach to reduce variance and stabilize kNN-based LID estimation on a number of utterances across different speakers. We further restrict this set to the intersection of utterances present across all perturbations in Table \ref{tab:perturb_params}, 
yielding 918 utterances that form a fully paired evaluation set. 
This design ensures that every LID comparison across perturbation types and SNR levels is computed over identical utterances. The utterance count also provides sufficient frame-level pooling (${\sim}$230k--460k embeddings per condition after convolutional 
subsampling) to stabilize kNN-based LID estimation across speakers.

For a fair comparison across perturbation types, we generate benign acoustic noise and adversarial examples for all target SNR levels (0--40~dB) summarized in Table~\ref{tab:perturb_params}: (1) Gaussian noise, (2) babble noise, (3) speech noise, and (4) Adversarial PGD examples. 

PGD is run for 300 iterations to maximize the MSE or CTC objective, rescaling the final projected perturbation to fully exhaust the per-utterance SNR budget $\epsilon_{\mathrm{SNR}}$. In the current kNN pipeline, the neighborhood size is selected by a per-condition $k$-sweep rather than assumed fixed a priori: candidate $k$ values are ranked by overall $\Delta$LID, filtered to those near the best value, and the final choice is taken to be the one with minimum across-layer $\Delta$LID variability. Section~\ref{subsec:lid_estimation_and_k} summarizes the resulting robustness behavior and also includes a compact fixed-$k$ comparison for reference.

We select WavLM and wav2vec~2.0 \textsc{Base} models ($L=12$, $d=768$) as complementary test cases representing distinct pretraining strategies. Both are trained on LibriSpeech 960h, but WavLM additionally incorporates a masked speech denoising objective that exposes the model to simulated overlapped speech during pretraining~\cite{Chen2021WavLMLS}, whereas wav2vec~2.0 is trained with a contrastive objective on clean speech only~\cite{baevski2020wav2vec}. This pairing allows us to assess whether denoising-aware pretraining is reflected in the geometric robustness of learned representations under perturbation.

Our \textit{GRIDS} framework is not inherently tied to a specific downstream task: LID operates directly on learned representations independently of task-specific labels or outputs. To evaluate LID as a diagnostic for ASR degradation, we compute WER using JiWER\footnote{https://jitsi.github.io/jiwer}. To quantify perturbation effectiveness, we apply Eq.~\eqref{eq:success_rate_and_wer} with $(\gamma,\tau)=(0.2,\,0.3)$ such that a successful attack both increases WER by at least $\gamma$ and produces a WER of at least $\tau$.

For anomaly detection, evaluation uses 5-fold grouped cross-validation over normalized utterance IDs rather than plain random or stratified folds, ensuring that all perturbation variants of the same underlying utterance remain within a single fold. This grouped protocol is used when reporting AUROC, AUPRC, and FPR@0.95 in Section~\ref{subsec:anomaly_detection}.

For ASR evaluation, we use the pretrained ASR models associated with wav2vec~2.0\footnote{https://huggingface.co/facebook/wav2vec2-base-960h} and WavLM\footnote{https://huggingface.co/patrickvonplaten/wavlm-libri-clean-100h-base}. The encoder and CTC head are frozen, and decoding uses greedy decoding. The same decoding configuration is used for both clean and perturbed utterances, and WER is computed from the decoded transcripts using JiWER.

\newcommand{\modelnorm}[2]{#1$_{\text{#2}}$}
\newcommand{\modelbold}[2]{\textbf{#1$_{\text{#2}}$}}

\section{Results and Analysis}\label{sec:results}
We report results for the following three analyses aligned to our \textit{GRIDS} framework: (i)~\textit{LID-S3M geometric analysis} for layer-wise LID under benign and adversarial perturbations; (ii)~\textit{LID-ASR monitoring} through empirical evidence that supports the co-occurrence of $\Delta$LID--WER under a range of target SNRs and perturbation types; and (iii)~\textit{LID-AD anomaly detection} and classifier performance results trained with our 12 LID-derived features from learned representations in WavLM and wav2vec~2.0. 

\subsection{LID-S3M Geometric Analysis}
\subsubsection{LID Estimation and \texorpdfstring{$k$}{k}-Sensitivity}\label{subsec:lid_estimation_and_k} We pool frame-level embeddings across all utterances within each perturbation setting in Table \ref{tab:perturb_params}, and compute layer-wise LID from $k$NN distances on the pooled set. $k$ is selected per condition rather than fixed globally. For each candidate neighborhood size, we compute the overall perturbation shift $\Delta$LID from Eq.~\eqref{eq:delta_lid_overall_def}, retain the candidates whose $\Delta$LID lies within a fixed fraction of the best value, and then choose the one with minimum across-layer standard deviation of $\Delta$LID, breaking ties by larger $\Delta$LID and then smaller $k$. This selection rule favors neighborhoods that remain both discriminative and stable across layers. Table~\ref{tab:delta_lid_k_sensitivity} should therefore be read as a compact fixed-$k$ robustness check rather than the full selection procedure. Across both models, perturbation ordering is qualitatively stable between $k{=}50$ and $k{=}100$: adversarial conditions yield the largest $\Delta$LID at low SNR, benign conditions move toward zero as SNR increases, and changing $k$ primarily rescales magnitude rather than reversing the trend. 

\newcommand{\thdr}[2]{\shortstack{#1\\$k{=}#2$}}
\begin{table}[!ht]
\centering
\footnotesize
\setlength{\tabcolsep}{3pt}
\caption{Sensitivity of overall $\Delta$LID to the neighborhood size $k$, evaluated at $k\in\{50,100\}$ for SNR 0 and 40~dB, computed via Eqs.~\eqref{eq:delta_lid_def}--\eqref{eq:overall_lid_def}.}
\label{tab:delta_lid_k_sensitivity}
\begin{tabularx}{\columnwidth}{@{}l X S S !{\vrule width 0.4pt} S S@{}}
\toprule
\multirow{2}{*}{\textbf{Model}} & \multirow{2}{*}{\textbf{Perturbation}}
& \multicolumn{2}{c}{$\Delta LID$ (0~dB)}
& \multicolumn{2}{c}{$\Delta LID$ (40~dB)} \\
\cmidrule(lr){3-4}\cmidrule(lr){5-6}
& & {$k{=}50$} & {$k{=}100$} & {$k{=}50$} & {$k{=}100$} \\
\midrule

\multirow{5}{*}{\textbf{WavLM}}
& PGD-MSE         & 12.40 & 12.46 & 2.24 & 2.10 \\
& PGD-CTC         & 11.30 & 11.46 & 1.99 & 1.84 \\
& Gaussian noise  &  2.51 &  2.35 & 0.79 & 0.60 \\
& Babble noise    &  2.99 &  4.48 & 0.84 & 0.67 \\
& Speech noise    &  4.85 &  5.52 & 0.76 & 0.64 \\
\midrule

\multirow{5}{*}{\textbf{wav2vec~2.0}}
& PGD-MSE         &  9.62 &  9.42 & 1.66 & 1.95 \\
& PGD-CTC         &  9.57 & 10.60 & 0.55 & 0.69 \\
& Gaussian noise  &  1.73 &  1.94 & 0.09 & 0.08 \\
& Babble noise    &  5.88 &  7.51 & 0.35 & 0.37 \\
& Speech noise    &  5.18 &  6.94 & 0.39 & 0.43 \\
\bottomrule
\end{tabularx}
\end{table}

\begin{figure*}[t]
\centering
\vspace{-1.2em}
\captionsetup{font=small}
\begin{subfigure}[t]{0.43\textwidth}
  \centering
  \includegraphics[width=\linewidth, trim=0cm 0cm 0cm 1.8cm, clip]{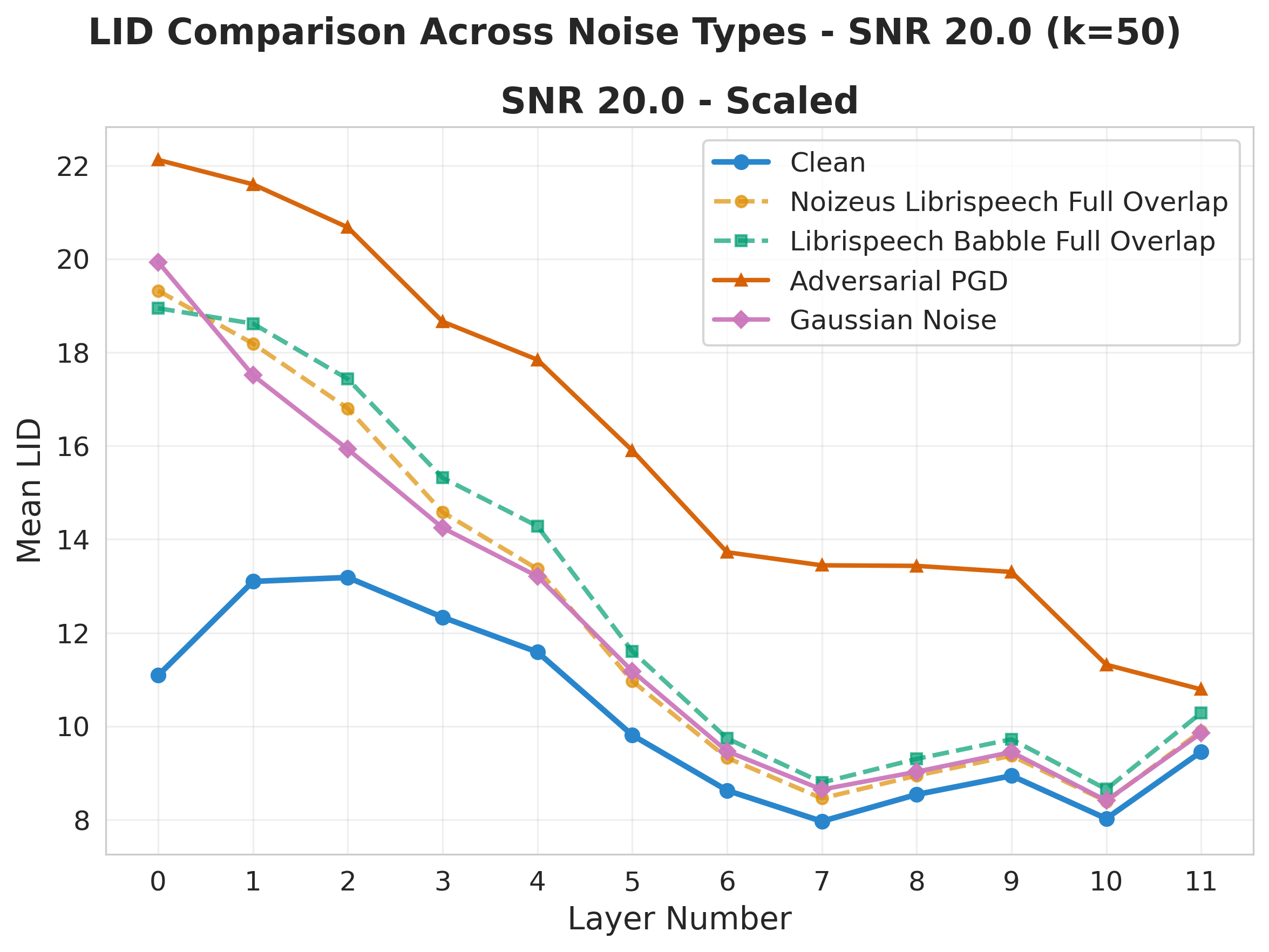}
  \caption{SNR 20~dB.}
  \label{fig:lid_mean_snr20_k50_mse}
\end{subfigure}\hspace{1em}
\begin{subfigure}[t]{0.43\textwidth}
  \centering
  \includegraphics[width=\linewidth, trim=0cm 0cm 0cm 1.85cm, clip]{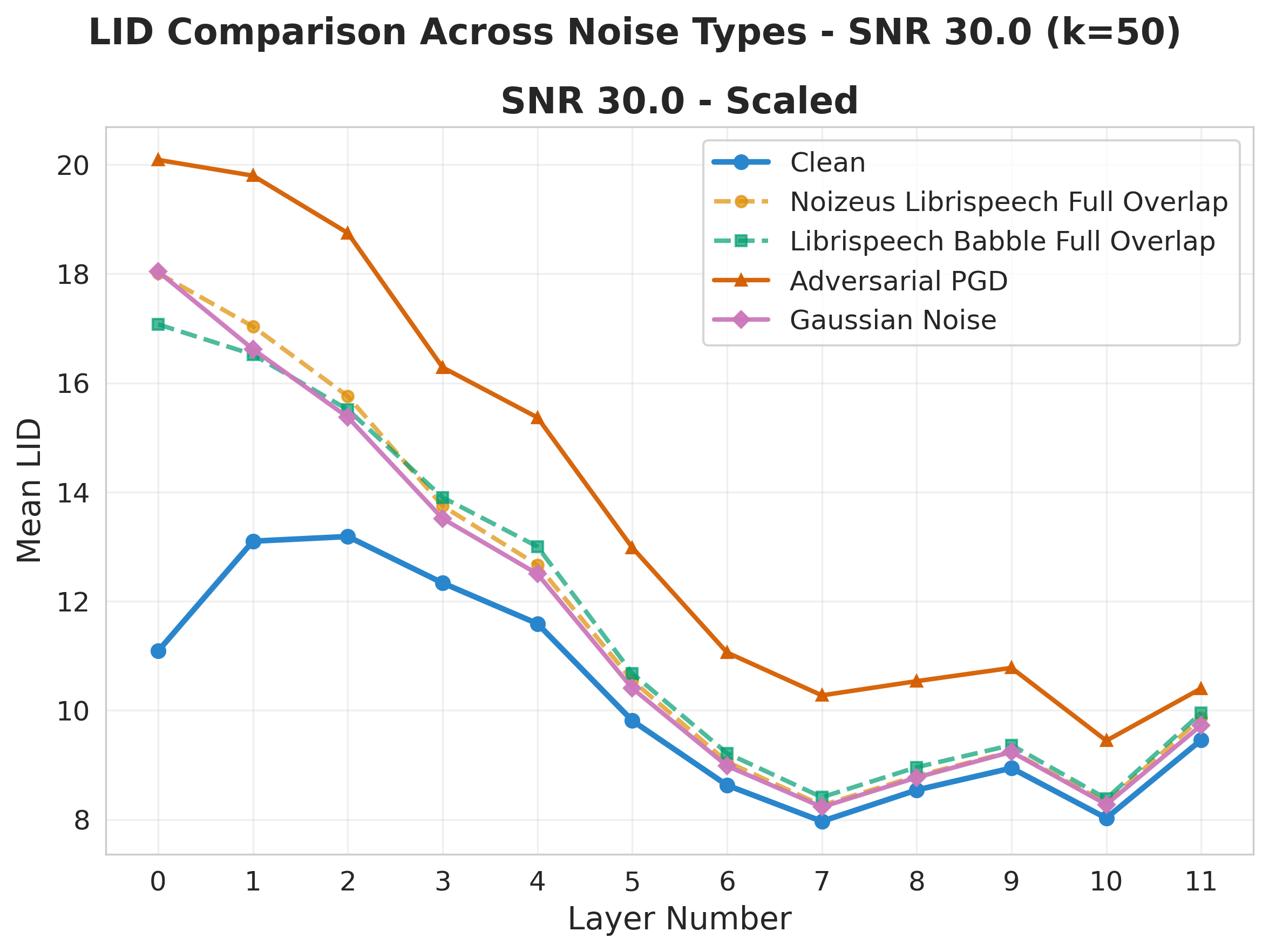}
  \caption{SNR 30~dB.}
  \label{fig:lid_mean_snr30_k50_mse}
\end{subfigure}\hspace{1em}
\captionsetup{list=no}
\caption{Layer-wise harmonic mean LID under \textbf{MSE-PGD} for WavLM at SNR 20/30~dB (k=50)} 
\label{fig:lid_mean_snr_sweep_k50_mse_wavlm}
\end{figure*}

\begin{figure*}[t]
\centering
\vspace{-1.2em}
\captionsetup{font=small}
\setlength{\tabcolsep}{3pt}

\begin{subfigure}[t]{0.43\textwidth}
  \centering
  \includegraphics[width=\linewidth, trim=0cm 0cm 0cm 1.8cm, clip]{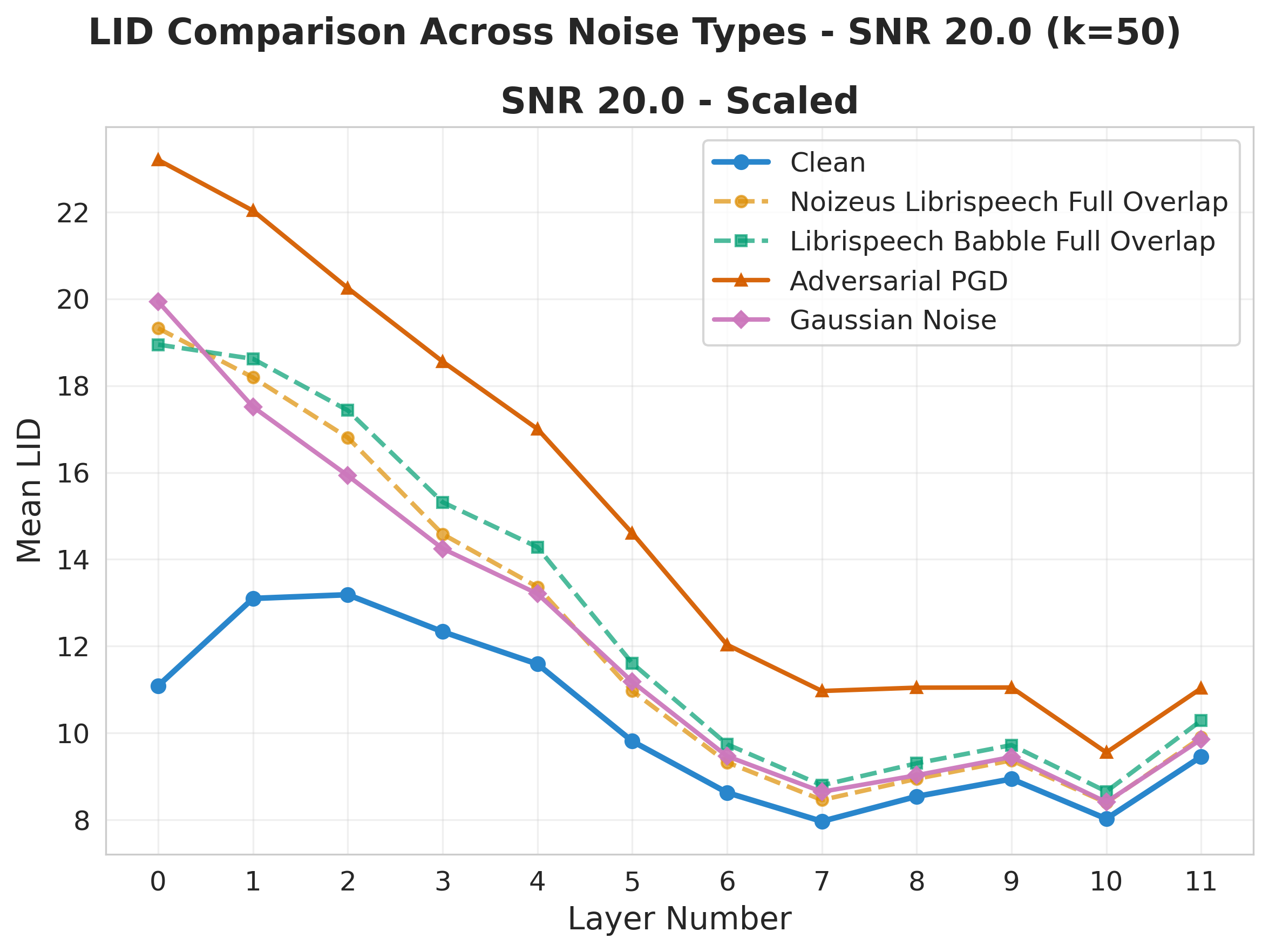}
  \caption{SNR 20~dB.}
  \label{fig:lid_mean_snr20_k50_ctc_wavlm}
\end{subfigure}\hspace{1em}
\begin{subfigure}[t]{0.43\textwidth}
  \centering
  \includegraphics[width=\linewidth, trim=0cm 0cm 0cm 1.8cm, clip]{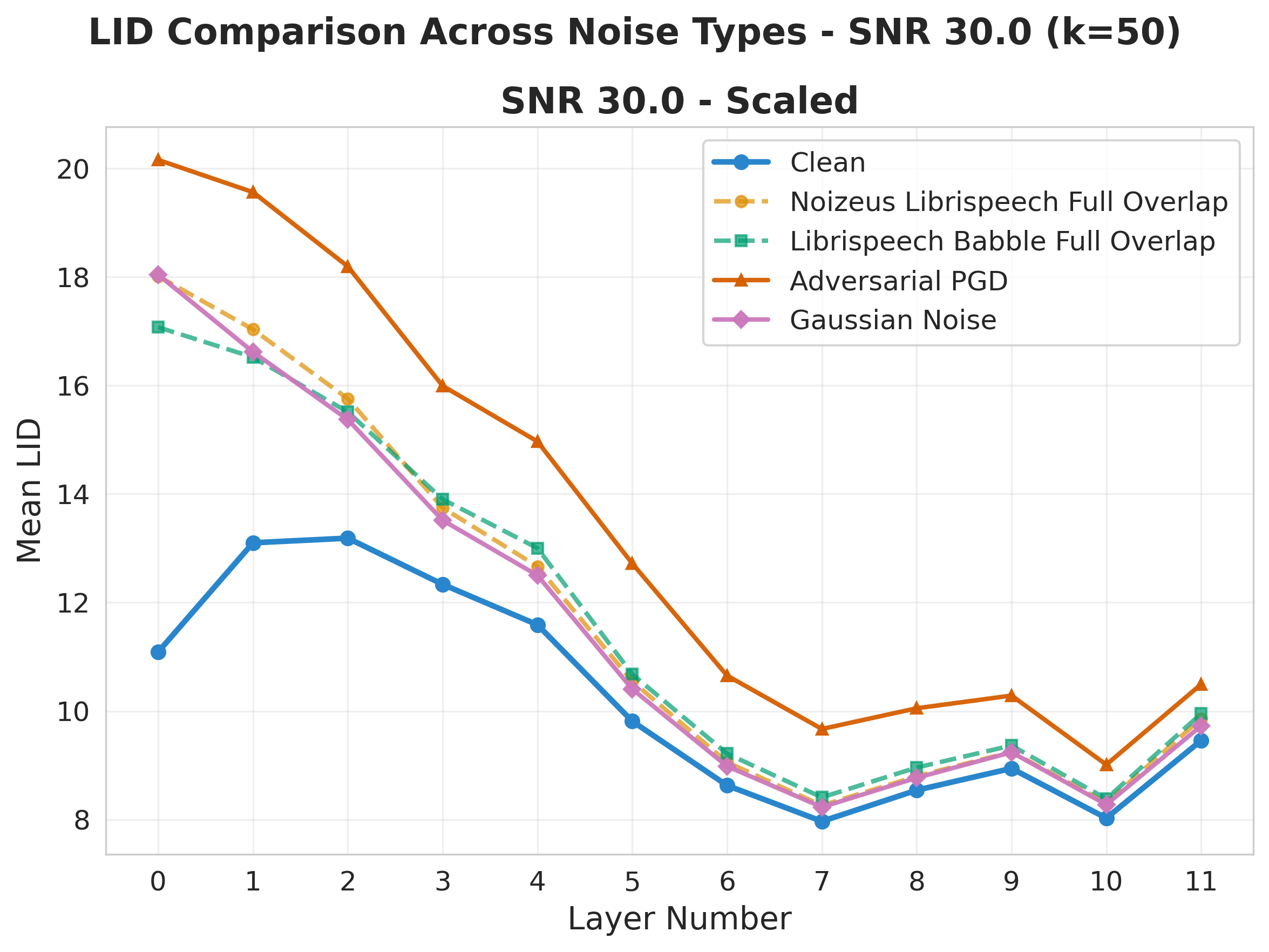}
  \caption{SNR 30~dB.}
  \label{fig:lid_mean_snr30_k50_ctc_wavlm}
\end{subfigure}

\caption{Layer-wise harmonic mean LID under \textbf{CTC-PGD} for WavLM at SNR 20/30~dB (k=50)}
\label{fig:lid_mean_snr_sweep_k50_ctc_wavlm}
\end{figure*}

\subsubsection{Layer-wise LID Trajectories}\label{subsec:lid_layerwise_trajectories}
We analyze how hidden-space geometry evolves across transformer layers by computing a layer-wise LID for clean speech and all perturbations in Table~\ref{tab:perturb_params}. We restrict benign speech-like interference to full-temporal overlap so that corruption spans all frames, yielding a conservative naturalistic baseline that is directly comparable to SNR-constrained adversarial perturbations. Across conditions, the mean LID typically decreases from early to late layers, coherent with progressive representational compression, from low-level acoustics to higher-level abstractions as noted in previous work \cite{pasad2023comparative} capturing the layer-wise trajectories in S3Ms. 

In our experiments, perturbations generally appear as upward shifts in LID across layers, relative to clean speech under a target SNR. We interpret higher LID as evidence of less compact or more locally complex neighborhood structure in the learned representation space, not as a direct measurement of global manifold dimension. This makes layer-wise LID useful for comparing how different perturbation families deform local geometry at the same layer, while the direction and persistence of those shifts are analyzed in the following section.

\subsubsection{Layer-wise LID in WavLM and wav2vec~2.0}
Figures \ref{fig:lid_mean_snr_sweep_k50_mse_wavlm}-\ref{fig:lid_mean_snr_sweep_k50_ctc_wavlm} show that adversarial perturbations are typically associated with the largest and most persistent LID elevation in WavLM. At low target SNR (0--10 dB), however, both babble and overlaid speech also increase LID substantially, so elevated LID should not be treated as uniquely adversarial. As target SNR increases to 30--40 dB, benign curves move closer to the clean profile, especially in later layers, while adversarial conditions more often retain elevated early-layer LID. We therefore interpret early-layer persistence as a relative marker of adversarial geometric shift under matched target-SNR conditions, rather than as a binary signature that is entirely absent from benign interference.

\begin{figure*}[t]
\centering
\vspace{-1.2em}
\captionsetup{font=small}
\setlength{\tabcolsep}{3pt}

\begin{subfigure}[t]{0.43\textwidth}
  \centering
  \includegraphics[width=\linewidth, trim=0cm 0cm 0cm 1.9cm, clip]{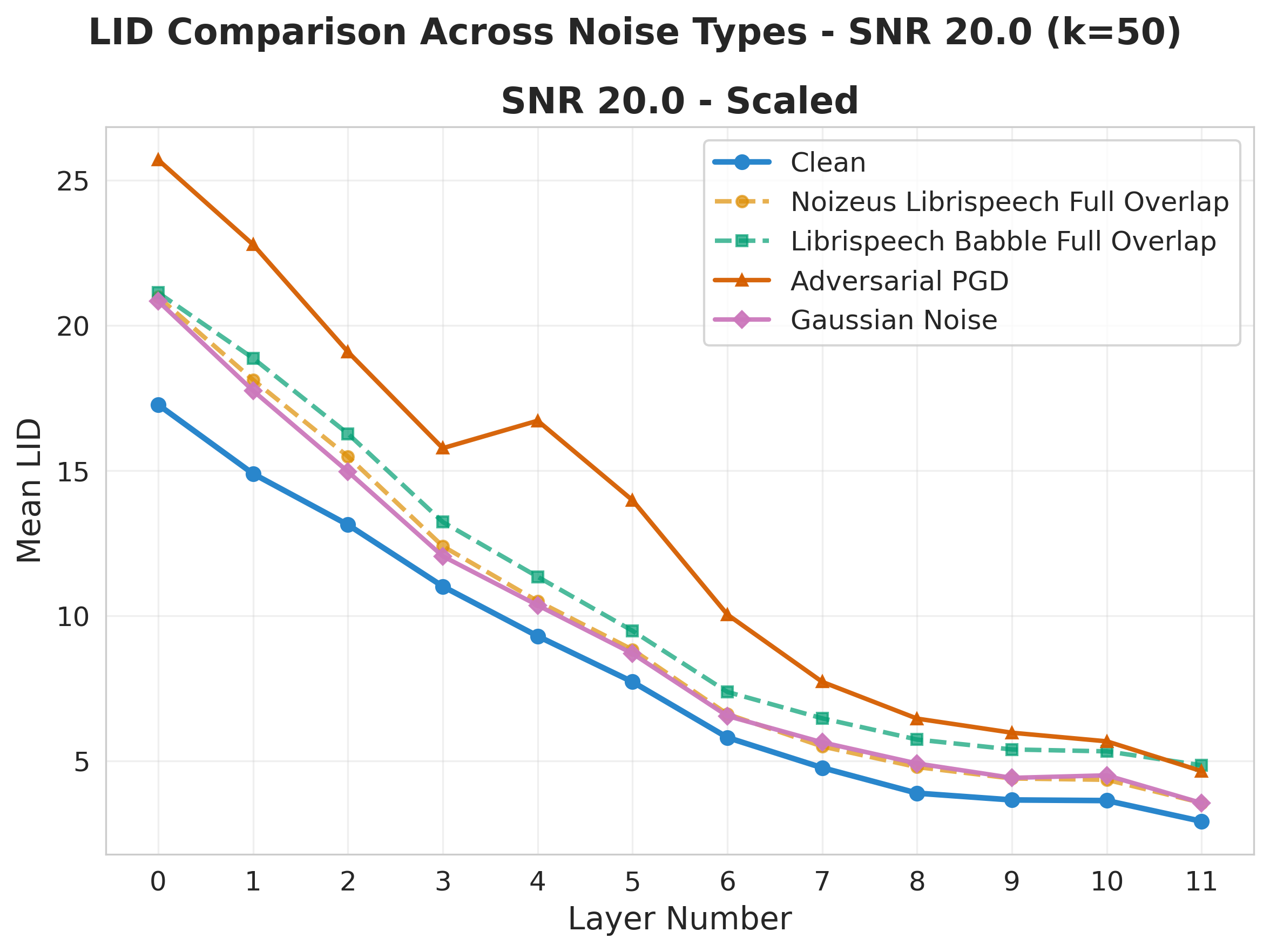}
  \caption{SNR 20~dB.}
  \label{fig:lid_mean_snr20_k50_mse_w2v2}
\end{subfigure}\hspace{1em}
\begin{subfigure}[t]{0.43\textwidth}
  \centering
  \includegraphics[width=\linewidth, trim=0cm 0cm 0cm 1.85cm, clip]{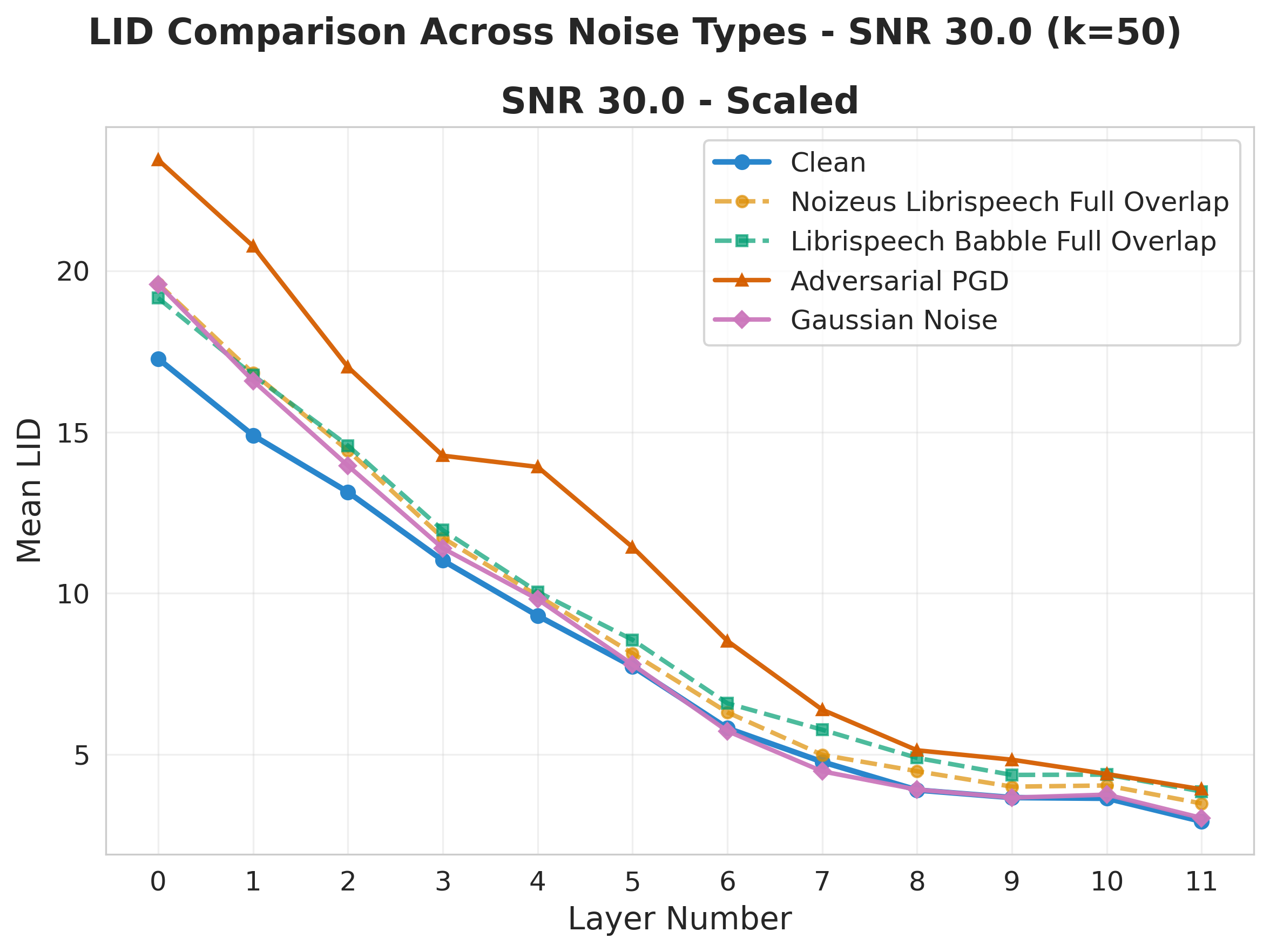}
  \caption{SNR 30~dB.}
  \label{fig:lid_mean_snr30_k50_mse_w2v2}
\end{subfigure}

\caption{Layer-wise harmonic mean LID under \textbf{MSE-PGD} for wav2vec~2.0 at SNR 20/30~dB (k=50).}
\label{fig:lid_mean_snr_sweep_k50_mse_w2v2}
\end{figure*}

\begin{figure*}[t]
\centering
\vspace{-1.2em}
\captionsetup{font=small}
\setlength{\tabcolsep}{3pt}

\begin{subfigure}[t]{0.43\textwidth}
  \centering
  \includegraphics[width=\linewidth, trim=0cm 0cm 0cm 1.95cm, clip]{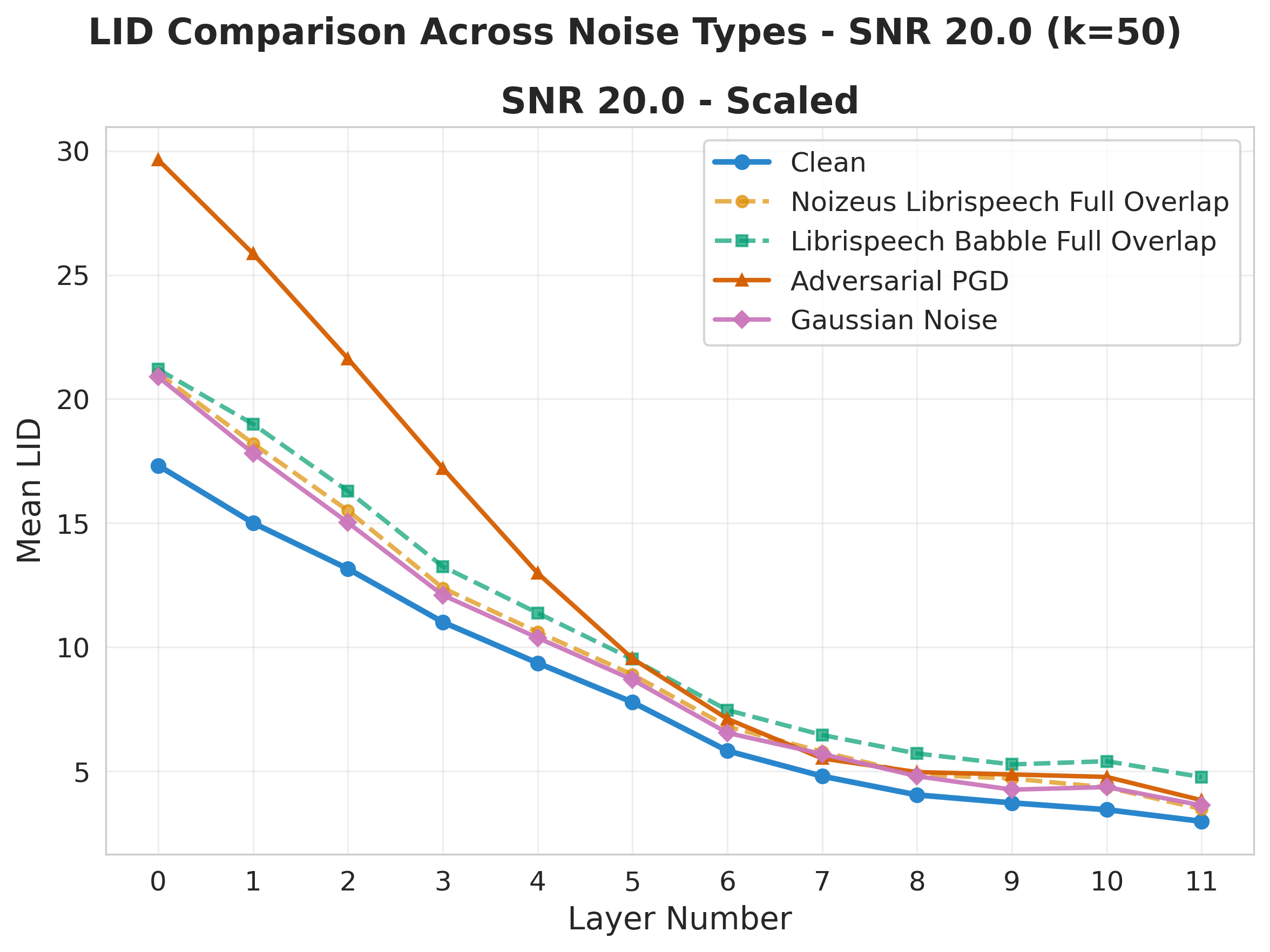}
  \caption{SNR 20~dB.}
  \label{fig:lid_mean_snr20_k50_ctc_w2v2}
\end{subfigure}\hspace{1em}
\begin{subfigure}[t]{0.43\textwidth}
  \centering
  \includegraphics[width=\linewidth, trim=0cm 0cm 0cm 1.8cm, clip]{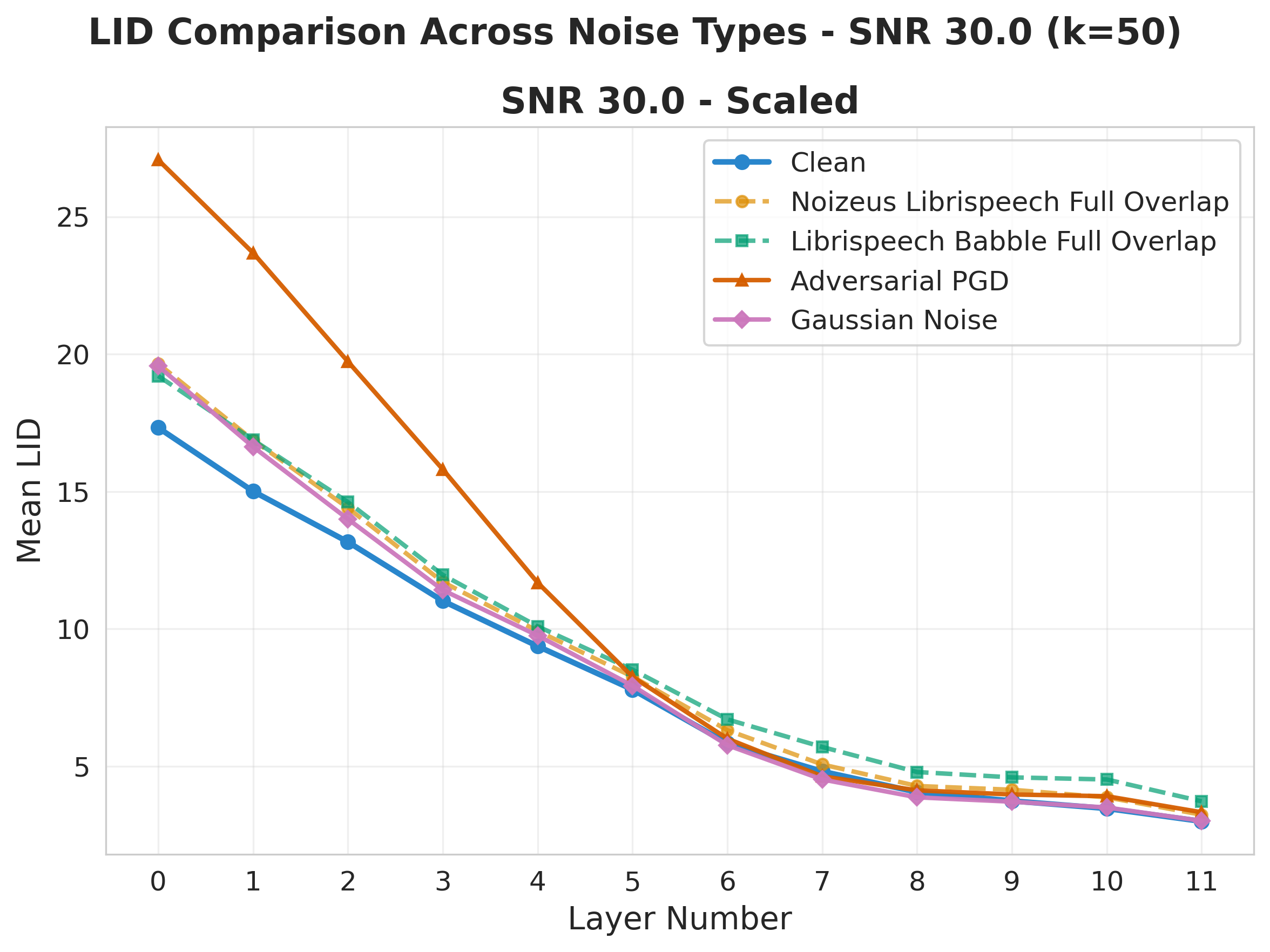}
  \caption{SNR 30~dB.}
  \label{fig:lid_mean_snr30_k50_ctc_w2v2}
\end{subfigure}
\caption{Layer-wise harmonic mean LID under \textbf{CTC-PGD} for wav2vec~2.0 at SNR 20/30~dB (k=50).}
\label{fig:lid_mean_snr_sweep_k50_ctc_w2v2}
\end{figure*}

To test whether this behavior transfers across architectures, we repeat the analysis for wav2vec~2.0 under the same perturbation families and target-SNR settings used for WavLM. Figures \ref{fig:lid_mean_snr_sweep_k50_mse_w2v2}-\ref{fig:lid_mean_snr_sweep_k50_ctc_w2v2} show the same broad pattern: adversarial conditions remain above benign distortions more often than not, the largest separations occur in early layers, and benign perturbations converge more rapidly toward the clean profile as target SNR increases. wav2vec~2.0 also exhibits a broad mid-layer peak, indicating that the precise layer profile is model-dependent even when the relative ordering across perturbation types is similar. We therefore treat elevated LID under adversarial perturbations as a cross-model tendency rather than as a WavLM-specific artifact.

\subsubsection{S3M robustness under varying SNR} 
Figures \ref{fig:lid_mean_snr_sweep_k50_mse_wavlm}-\ref{fig:lid_mean_snr_sweep_k50_ctc_w2v2} show that the degree and shape of LID persistence depends on the attack objective (PGD-MSE vs. task-aligned PGD-CTC), with PGD-MSE producing the largest and most persistent elevation under the same SNR budget. The clearest separation across perturbation types occurs in the early transformer layers, suggesting that layer-wise LID provides a more informative robustness signal than a single final-layer summary.

\subsection{LID-ASR monitoring}\label{subsec:wer_lid_corr}

\subsubsection{LID and WER Analysis}
\newcommand{\lw}[2]{#1\,/\,\textbf{#2}}
\begin{table}[!hb]
\centering
\scriptsize
\caption{Per-SNR overall $\Delta$LID and $\Delta$WER on the paired 918-utterance LibriSpeech \texttt{test-clean} subset. Clean baseline WER is 0.04 for both models.}
\setlength{\tabcolsep}{4pt}
\renewcommand{\arraystretch}{1.1}
\begin{tabular}{c c c c c c}
\toprule
\textbf{SNR} & \textbf{PGD-CTC} & \textbf{PGD-MSE} & \textbf{Gaussian} & \textbf{Babble} & \textbf{Speech} \\
\textbf{(dB)} & \multicolumn{5}{c}{$\Delta$LID\,/\,\textbf{$\Delta$WER}} \\
\midrule
\multicolumn{6}{l}{\textbf{WavLM}} \\
0  & \lw{10.51}{0.84} & \lw{16.03}{0.94} & \lw{1.67}{0.04} & \lw{3.50}{0.72} & \lw{4.63}{1.00} \\
10 & \lw{5.71}{0.46}  & \lw{12.02}{0.83} & \lw{1.67}{0.04} & \lw{1.88}{0.11} & \lw{2.98}{0.46} \\
20 & \lw{3.42}{0.20}  & \lw{6.71}{0.47}  & \lw{1.10}{0.01} & \lw{0.92}{0.01} & \lw{1.49}{0.10} \\
30 & \lw{2.23}{0.08}  & \lw{2.23}{0.14}  & \lw{0.55}{0.00} & \lw{0.63}{0.00} & \lw{0.77}{0.01} \\
40 & \lw{1.42}{0.03}  & \lw{1.71}{0.04}  & \lw{0.36}{0.00} & \lw{0.45}{0.00} & \lw{0.43}{0.00} \\
\midrule
\multicolumn{6}{l}{\textbf{wav2vec~2.0}} \\
0  & \lw{6.21}{0.76}  & \lw{10.11}{0.96} & \lw{1.35}{0.04} & \lw{7.51}{0.72} & \lw{5.79}{1.00} \\
10 & \lw{3.06}{0.23}  & \lw{5.74}{0.90}  & \lw{1.37}{0.04} & \lw{4.15}{0.11} & \lw{4.74}{0.46} \\
20 & \lw{1.43}{0.05}  & \lw{2.70}{0.51}  & \lw{1.00}{0.01} & \lw{1.36}{0.01} & \lw{2.74}{0.10} \\
30 & \lw{0.27}{0.02}  & \lw{1.19}{0.10}  & \lw{0.13}{0.00} & \lw{0.60}{0.00} & \lw{0.96}{0.01} \\
40 & \lw{0.00}{0.01}  & \lw{0.76}{0.01}  & \lw{0.00}{0.00} & \lw{0.22}{0.00} & \lw{0.37}{0.00} \\
\bottomrule
\end{tabular}
\label{tab:wer_lid_snr_compact}
\end{table}

As noted in Section~\ref{sec:intro}, if geometric structure is integral to how S3Ms encode information across layers, then perturbations that disrupt representation geometry may also be associated with downstream performance degradation. We test this by examining whether the magnitude of LID-derived geometric shifts tracks the degree of ASR degradation across perturbation types and target-SNR conditions.

Rather than reporting WER for adversarial PGD in isolation, we summarize WER jointly with $\Delta$LID for \emph{all} perturbation types under the layer-wise LID analysis detailed in Section~\ref{subsec:lid_layerwise_trajectories}. For each condition, we compute two ASR metrics from per-utterance transcripts: $\mathrm{WER}_{\mathrm{pert}}$ is the arithmetic mean of per-utterance WER values on perturbed audio, and $\Delta\mathrm{WER}$ is the mean of per-utterance differences, i.e.,
\[
\Delta\mathrm{WER}
=
\mathbb{E}_i\ \left[
\mathrm{WER}_{\mathrm{pert},i}-\mathrm{WER}_{\mathrm{clean},i}
\right].
\]

Table~\ref{tab:wer_lid_snr_compact} reports $\Delta\mathrm{LID}$ using Eq.~\eqref{eq:delta_lid_overall_def} to reflect geometric shift, alongside $\Delta$WER, showing a consistent co-occurrence between geometric shift and ASR degradation under matched SNR: both decrease as SNR rises from 0 to 40~dB across conditions. The association is most apparent at low SNR (0--10~dB), where adversarial PGD and speech-like interference (babble and overlaid speech) jointly produce the largest deviations from clean behavior. At 0~dB, PGD-MSE induces the largest geometric deformation ($\Delta$LID 16.03 for WavLM, 10.11 for wav2vec~2.0), accompanied by large error increases ($\Delta$WER above 0.94). Overlaid speech maximizes $\Delta$WER (1.00 for both models) at more moderate $\Delta$LID (4.63 and 5.79), showing that severe transcription failure can occur even when the geometric shift is not maximal. Gaussian noise remains low-impact throughout ($\Delta$WER$\le$0.04). Within a fixed SNR, adversarial conditions typically dominate both axes relative to benign noise at comparable energy, and PGD-MSE is more disruptive than PGD-CTC. At high SNR (30--40~dB), benign corruptions become nearly clean-like in $\Delta$WER, whereas PGD retains a non-trivial $\Delta$LID footprint, indicating residual geometric displacement despite small decoding error.

\subsection{Anomaly Detection}\label{subsec:anomaly_detection}
Using the 12-dimensional LID feature vector defined in Eq.~\eqref{eq:lid_feature_vector}, we train a lightweight logistic-regression classifier per SNR, with scores evaluated out-of-fold under grouped 5-fold cross-validation. Detection performance is overall strong but varies with both model and attack objective: \modelnorm{WavLM}{MSE} is the most separable configuration (mean AUROC 0.99, mean FPR@0.95 = 0.02), while \modelnorm{wav2vec2}{MSE} is the least (mean AUROC 0.88, mean FPR@0.95 = 0.39). The relative difficulty of MSE- versus CTC-PGD is model-dependent: MSE is more separable than CTC for WavLM, whereas the ordering reverses for wav2vec~2.0 (Table~\ref{tab:rq3_transposed_three_row}).

\newcommand{\fprsr}[2]{#1{\tiny[#2]}}
\begin{table}[!ht]
\centering
\scriptsize
\caption{Anomaly-detection performance for adversarial vs benign, evaluated with 5-fold GroupKFold. AUROC, AUPRC, FPR  at TPR$=0.95$, and success rate [SR]. wav2vec~2.0 is abbreviated as w2v2.}
\setlength{\tabcolsep}{2pt}
\renewcommand{\arraystretch}{1.05}
\begin{tabular}{@{}c l l l l l@{}}
\toprule
\textbf{SNR} & \textbf{METRIC} & \textbf{WavLM$_{\text{CTC}}$} & \textbf{WavLM$_{\text{MSE}}$} & \textbf{w2v2$_{\text{CTC}}$} & \textbf{w2v2$_{\text{MSE}}$} \\
\midrule
\multirow{3}{*}{0}
& AUROC      & 1.00 & 1.00 & 1.00 & 1.00 \\
& AUPRC      & 1.00 & 1.00 & 1.00 & 1.00 \\
& FPR@0.95[SR] & \fprsr{0.00}{1.00} & \fprsr{0.00}{1.00} & \fprsr{0.00}{1.00} & \fprsr{0.00}{1.00} \\
\midrule
\multirow{3}{*}{10}
& AUROC      & 1.00 & 1.00 & 1.00 & 0.97 \\
& AUPRC      & 0.99 & 0.99 & 0.99 & 0.91 \\
& FPR@0.95[SR] & \fprsr{0.01}{0.80} & \fprsr{0.01}{0.99} & \fprsr{0.01}{1.00} & \fprsr{0.15}{0.33} \\
\midrule
\multirow{3}{*}{20}
& AUROC      & 0.97 & 1.00 & 1.00 & 0.85 \\
& AUPRC      & 0.93 & 0.99 & 0.99 & 0.59 \\
& FPR@0.95[SR] & \fprsr{0.20}{0.28} & \fprsr{0.00}{0.73} & \fprsr{0.02}{0.78} & \fprsr{0.48}{0.02} \\
\midrule
\multirow{3}{*}{30}
& AUROC      & 0.92 & 1.00 & 0.99 & 0.80 \\
& AUPRC      & 0.83 & 0.99 & 0.98 & 0.52 \\
& FPR@0.95[SR] & \fprsr{0.43}{0.04} & \fprsr{0.00}{0.18} & \fprsr{0.05}{0.10} & \fprsr{0.62}{0.00} \\
\midrule
\multirow{3}{*}{40}
& AUROC      & 0.87 & 0.98 & 0.94 & 0.78 \\
& AUPRC      & 0.77 & 0.96 & 0.87 & 0.50 \\
& FPR@0.95[SR] & \fprsr{0.60}{0.01} & \fprsr{0.11}{0.03} & \fprsr{0.33}{0.00} & \fprsr{0.69}{0.00} \\
\bottomrule
\end{tabular}
\label{tab:rq3_transposed_three_row}
\end{table}
As SNR increases, the attack success rate (SR) defined in Eq.~\eqref{eq:success_rate_and_wer} drops sharply, consistent with the smaller $\ell_2$ perturbation budget at higher SNR. This decline in SR coincides with a general increase in FPR@0.95 and a decrease in AUROC, suggesting that reduced separability at high SNR is partly driven by a growing fraction of low-impact adversarial examples that fail to meaningfully degrade ASR. SR alone does not fully account for classifier performance: \modelnorm{WavLM}{CTC} achieves AUROC~0.92 at 30~dB despite SR of only 0.04, indicating that LID can retain discriminative geometric information even when WER does not exceed the success threshold.

\section{Conclusion}\label{sec:conclusion}
We have shown that layer-wise LID is an effective diagnostic for local geometric changes in WavLM and wav2vec~2.0 under benign and adversarial perturbations. Across both models, and despite distinct pretraining objectives, our analysis reveals that the clearest divergence between adversarial and benign profiles occurs in early transformer layers, suggesting geometric distortion propagates through early representations before being partially absorbed by deeper layers. We further demonstrate that shifts in representational geometry can be measured using LID and be linked to ASR degradation without requiring ground-truth transcripts: higher LID co-occurs with WER increases. A lightweight logistic-regression classifier on our 12 LID-derived features distinguishes adversarial from benign inputs across SNRs, with separability decreasing at high SNR as both converge toward the clean manifold. Our \textit{GRIDS} framework thus contributes an interpretable geometric diagnostic for transcript-free monitoring in S3Ms. Current limitations include restriction to 12-layer S3M variants and untargeted attacks. Future work includes scaling to larger architectures, extending to security-critical tasks such as speaker verification and emotion recognition, and combining LID with global spectral measures (effective rank~\cite{aldeneh2024rankme,rank-Whetten2025}, PCA variance decomposition~\cite{gurnee2026modelsmanipulatemanifoldsgeometry_anthropic}) for multi-scale geometric characterization.

\section{Acknowledgments}
This research was supported by the Australian Government Research Training Program Scholarship [DOI: https://doi.org/10.82133/C42F-K220].

\section{Declaration on Generative AI}
The author(s) used ChatGPT and Claude to edit, check grammar, spelling, and minor paraphrasing. All technical claims, metrics, and artifact references were manually verified.

\bibliographystyle{IEEEtran}
\bibliography{mybib}

@article{Mohamed2022-ssl-review,
    title={Self-supervised speech representation learning: A review},
    author={Mohamed, Abdelrahman and Lee, Hung-yi and Borgholt, Lasse and Havtorn, Jakob D and Edin, Joakim and Igel, Christian and Kirchhoff, Katrin and Li, Shang-Wen and Livescu, Karen and Maal{\o}e, Lars and others},
    journal={IEEE Journal of Selected Topics in Signal Processing},
    volume={16},
    number={6},
    pages={1179--1210},
    year={2022},
    publisher={IEEE}
}

@article{Hsu_2021-HuBERT,
    author = {Hsu, Wei-Ning and Bolte, Benjamin and Tsai, Yao-Hung Hubert and Lakhotia, Kushal and Salakhutdinov, Ruslan and Mohamed, Abdelrahman},
    title = {HuBERT: Self-Supervised Speech Representation Learning by Masked Prediction of Hidden Units},
    year = {2021},
    issue_date = {2021},
    publisher = {IEEE Press},
    volume = {29},
    issn = {2329-9290},
    journal = {IEEE/ACM Trans. Audio, Speech and Lang. Proc.},
    pages = {3451--3460}
}

@inproceedings{kornblith2019similarity-cka-Hinton,
    title = 	 {Similarity of Neural Network Representations Revisited},
    author =       {Kornblith, Simon and Norouzi, Mohammad and Lee, Honglak and Hinton, Geoffrey},
    booktitle = 	 {Proc. ICML},
    pages = 	 {3519--3529},
    year = 	 {2019},
    publisher =    {PMLR},
}

@inproceedings{Pasad_2021-cca-layerwise-Livescu,
    author={Pasad, Ankita and Chou, Ju-Chieh and Livescu, Karen},
    booktitle={Proc. ASRU}, 
    title={Layer-Wise Analysis of a Self-Supervised Speech Representation Model}, 
    publisher = {IEEE},
    year={2021},
    pages={914-921},
    doi={10.1109/ASRU51503.2021.9688093}
  }

@inproceedings{huang2025detectingbackdoorsamplescontrastive-localsubspace-LID-JamesB,
    author = {Hanxun Huang and Sarah M. Erfani and Yige Li and Xingjun Ma and James Bailey},
    title={Detecting Backdoor Samples in Contrastive Language Image Pretraining},
    booktitle={Proc. ICLR},
    year={2025},
}

@inproceedings{raghu2017svcca-brno-CKA-CCA_CNN-LSTM-transformer,
    author = {Raghu, Maithra and Gilmer, Justin and Yosinski, Jason and Sohl-Dickstein, Jascha},
    title = {SVCCA: singular vector canonical correlation analysis for deep learning dynamics and interpretability},
    year = {2017},
    booktitle = {Proc. NeurIPS},
    pages = {6078--6087},
}

@inproceedings{baevski2020wav2vec,
    author = {Baevski, Alexei and Zhou, Henry and Mohamed, Abdelrahman and Auli, Michael},
    title = {wav2vec 2.0: a framework for self-supervised learning of speech representations},
    year = {2020},
    isbn = {9781713829546},
    publisher = {Curran Associates Inc.},
    booktitle = {Proc. NeurIPS},
    articleno = {1044}
}

@article{Chen2021WavLMLS,
    author={Chen, Sanyuan and Wang, Chengyi and Chen, Zhengyang and Wu, Yu and Liu, Shujie and Chen, Zhuo and Li, Jinyu and Kanda, Naoyuki and Yoshioka, Takuya and Xiao, Xiong and Wu, Jian and Zhou, Long and Ren, Shuo and Qian, Yanmin and Qian, Yao and Wu, Jian and Zeng, Michael and Yu, Xiangzhan and Wei, Furu},
    journal={IEEE Journal of Selected Topics in Signal Processing}, 
    title={WavLM: Large-Scale Self-Supervised Pre-Training for Full Stack Speech Processing}, 
    year={2022},
    volume={16},
    number={6},
    pages={1505-1518},
    doi={10.1109/JSTSP.2022.3188113}
}

@inproceedings{ma2018characterizing,
  title={Characterizing adversarial subspaces using local intrinsic dimensionality},
  author={Xingjun Ma and Bo Li and Yisen Wang and Sarah M. Erfani and Sudanthi Wijewickrema and Grant Schoenebeck and Dawn Song and Michael E. Houle and James Bailey},
  booktitle={Proc. ICLR},
  year={2018}
}

@inproceedings{morcos2018insights,
    author = {Morcos, Ari S. and Raghu, Maithra and Bengio, Samy}, 
    title = {Insights on representational similarity in neural networks with canonical correlation}, 
    year = {2018}, 
    booktitle = {Proc. NeurIPS}, 
    pages = {5732--5741}
}

@inproceedings{Houle2018,
    author = {Houle, Michael E. and Schubert, Erich and Zimek, Arthur}, title = {On the Correlation Between Local Intrinsic Dimensionality and Outlierness}, year = {2018}, 
    isbn = {978-3-030-02223-5}, 
    publisher = {Springer-Verlag}, 
    doi = {10.1007/978-3-030-02224-2_14},
    booktitle = {11th International Conference of Similarity Search and Applications}, 
    pages = {177--191}
}

@inproceedings{ruppik2025less,
    title={Less is More: Local Intrinsic Dimensions of Contextual Language Models},
    author={Benjamin Matthias Ruppik and Julius von Rohrscheidt and Carel van Niekerk and Michael Heck and Renato Vukovic and Shutong Feng and Hsien-chin Lin and Nurul Lubis and Bastian Rieck and Marcus Zibrowius and Milica Gasic},
    booktitle={Proc. NeurIPS},
    year={2025},
}

@inproceedings{NIPS2004_74934548,
     author = {Levina, Elizaveta and Bickel, Peter J.}, 
     title = {Maximum Likelihood estimation of intrinsic dimension}, year = {2004}, 
     publisher = {MIT Press}, 
     booktitle = {Proc. NeurIPS}, 
     pages = {777--784}
    }

@inproceedings{pasad2023comparative,
      author={Pasad, Ankita and Shi, Bowen and Livescu, Karen},
      booktitle={Proc. ICASSP}, 
      title={Comparative Layer-Wise Analysis of Self-Supervised Speech Models}, 
      year={2023},
      pages={1-5},
      doi={10.1109/ICASSP49357.2023.10096149}
}

@inproceedings{aldeneh2024rankme,
      author = {Zakaria Aldeneh and
                  Vimal Thilak and
                  Takuya Higuchi and
                  Barry{-}John Theobald and
                  Tatiana Likhomanenko},

    title = {Towards Automatic Assessment of Self-Supervised Speech Models using
                  Rank},
    booktitle    = {Proc. ICASSP},
    pages        = {1--5},
    publisher    = {{IEEE}},
    year         = {2025},
    doi          = {10.1109/ICASSP49660.2025.10889651}
}

@INPROCEEDINGS{rank-Whetten2025,
  title     = "Towards early prediction of self-supervised speech model
               performance",
  author    = "Whetten, Ryan and Maison, Lucas and Parcollet, Titouan and
               Dinarelli, Marco and Estève, Yannick",
  booktitle = "Proc. Interspeech",
  pages     = "1228--1232",
  year      =  2025
}

@ARTICLE{Esmaeilpour2019-adversarial-audio-dnns,
  author={Esmaeilpour, Mohammad and Cardinal, Patrick and Lameiras Koerich, Alessandro},
  journal={IEEE Transactions on Information Forensics and Security}, 
  title={A Robust Approach for Securing Audio Classification Against Adversarial Attacks}, 
  year={2020},
  volume={15},
  number={},
  pages={2147-2159},
  doi={10.1109/TIFS.2019.2956591}}

@article{Liu2018Voice,
    author = {Boquan Liu and Evan Polce and Jack Jiang},
    title ={Application of Local Intrinsic Dimension for Acoustical Analysis of Voice Signal Components},
    journal = {Annals of Otology, Rhinology \& Laryngology},
    volume = {127},
    number = {9},
    pages = {588-597},
    year = {2018},
    doi = {10.1177/0003489418780439},
}

@article{NoizeusDataset,
author = {Yi Hu and Philipos C. Loizou},
title = {Subjective comparison and evaluation of speech enhancement algorithms},
journal = {Speech Communication},
volume = {49},
number = {7},
pages = {588-601},
year = {2007},
note = {Special issue on Speech Enhancement},
issn = {0167-6393},
doi = {https://doi.org/10.1016/j.specom.2006.12.006},
}

@inproceedings{LibriSpeechDataset,
  author={Panayotov, Vassil and Chen, Guoguo and Povey, Daniel and Khudanpur, Sanjeev},
  booktitle={Proc. ICASSP}, 
  title={Librispeech: An ASR corpus based on public domain audio books}, 
  year={2015},
  pages={5206-5210},
  doi={10.1109/ICASSP.2015.7178964}}

@inproceedings{madry2019-PGD,
        author = {Aleksander Madry and
                      Aleksandar Makelov and
                      Ludwig Schmidt and
                      Dimitris Tsipras and
                      Adrian Vladu},
      title        = {Towards Deep Learning Models Resistant to Adversarial Attacks},
      booktitle    = {Proc. ICLR},
      year         = {2018},
      bibsource    = {dblp computer science bibliography, https://dblp.org}
}

@inproceedings{graves2006-ctc,
    author = {Graves, Alex and Fern\'{a}ndez, Santiago and Gomez, Faustino and Schmidhuber, J\"{u}rgen},
    title = {Connectionist temporal classification: labelling unsegmented sequence data with recurrent neural networks},
    year = {2006},
    isbn = {1595933832},
doi = {10.1145/1143844.1143891},
booktitle = {Proc. ICML},
pages = {369--376}
}

@inproceedings{hsu2021robust_wav2vec2,
    title     = {{Robust wav2vec 2.0: Analyzing Domain Shift in Self-Supervised Pre-Training}},
    author    = {Wei-Ning Hsu and Anuroop Sriram and Alexei Baevski and Tatiana Likhomanenko and Qiantong Xu and Vineel Pratap and Jacob Kahn and Ann Lee and Ronan Collobert and Gabriel Synnaeve and Michael Auli},
    year      = {2021},
    booktitle = {Proc. Interspeech},
    pages     = {721--725},
    doi       = {10.21437/Interspeech.2021-236},
    issn      = {2958-1796},
}

@inproceedings{huang2022distortion_domain_adapt,
    author       = {Kuan{-}Po Huang and
                  Yu{-}Kuan Fu and
                  Yu Zhang and
                  Hung{-}yi Lee},
    title        = {Improving Distortion Robustness of Self-supervised Speech Processing
                  Tasks with Domain Adaptation},
    booktitle    = {Proc. Interspeech},
    pages        = {2193--2197},
    year         = {2022},
    doi          = {10.21437/INTERSPEECH.2022-519},
    bibsource    = {dblp computer science bibliography, https://dblp.org}
}

@inproceedings{wang2022speech_reconstruction,
  author = {Wang, Heming and Qian, Yao and Wang, Xiaofei and Wang, Yiming and Wang, Chengyi and Liu, Shujie and Yoshioka, Takuya and Li, Jinyu and Wang, DeLiang},
  title        = {Improving Noise Robustness of Contrastive Speech Representation Learning with Speech Reconstruction},
  booktitle    = {Proc. ICASSP},
  pages        = {6062--6066},
  publisher    = {IEEE},
  year         = {2022},
  doi          = {10.1109/ICASSP43922.2022.9746220},
  bibsource    = {dblp computer science bibliography, https://dblp.org}
}

@inproceedings{wu2021adv_vulnerability_speech,
      author       = {Haibin Wu and
                      Bo Zheng and
                      Xu Li and
                      Xixin Wu and
                      Hung{-}Yi Lee and
                      Helen Meng},
      title        = {Characterizing the Adversarial Vulnerability of Speech self-Supervised
                      Learning},
      booktitle    = {Proc. {ICASSP}},
      pages        = {3164--3168},
      publisher    = {{IEEE}},
      year         = {2022},
      doi          = {10.1109/ICASSP43922.2022.9747242},
      bibsource    = {dblp computer science bibliography, https://dblp.org}
}

@misc{gurnee2026modelsmanipulatemanifoldsgeometry_anthropic,
      title={When Models Manipulate Manifolds: The Geometry of a Counting Task}, 
      author={Wes Gurnee and Emmanuel Ameisen and Isaac Kauvar and Julius Tarng and Adam Pearce and Chris Olah and Joshua Batson},
      year={2026},
      eprint={2601.04480},
      archivePrefix={arXiv},
      primaryClass={cs.LG},
      url={https://arxiv.org/abs/2601.04480}, 
}

@article{fefferman2016testing,
  title={Testing the manifold hypothesis},
  author={Fefferman, Charles and Mitter, Sanjoy and Narayanan, Hariharan},
  journal={Journal of the American Mathematical Society},
  volume={29},
  number={4},
  pages={983--1049},
  year={2016}
}

@inproceedings{NIPS2010_8a1e808b,
    author = {Narayanan, Hariharan and Mitter, Sanjoy},
    booktitle = {Proc. NeurIPS},
    editor = {J. Lafferty and C. Williams and J. Shawe-Taylor and R. Zemel and A. Culotta},
    title = {Sample Complexity of Testing the Manifold Hypothesis},
    volume = {23},
    year = {2010}
}

@inproceedings{cca-nonlinear1,
  title     = "Deep Canonical Correlation Analysis",
  author    = "Andrew, Galen and Arora, Raman and Bilmes, Jeff and Livescu, Karen",
  booktitle = "Proc. ICML",
  publisher = "PMLR",
  pages     = "1247--1255",
  year      =  2013
}

@inproceedings{cca-nonlinear2,
  title     = "On deep multi-view representation learning",
  author    = "Wang, Weiran and Arora, Raman and Livescu, Karen and Bilmes, Jeff",
  booktitle = "Proc. ICML",
  publisher = "PMLR",
  pages     = "1083--1092",
  year      =  2015
}

@inproceedings{cca-nonlinear3,
  title     = "Nonlinear feature extraction using generalized canonical
               correlation analysis",
  author    = "Melzer, Thomas and Reiter, Michael and Bischof, Horst",
  booktitle = "International Conference on Artificial Neural Networks",
  pages     = "353--360",
  year      =  2001,
  doi = {10.1007/3-540-44668-0_50}
}

@article{cca-nonlinear4,
  title     = "Kernel and nonlinear canonical correlation analysis",
  author    = "Lai, P L and Fyfe, C",
  journal   = "International Journal of Neural Systems",
  volume    =  10,
  number    =  5,
  pages     = "365--377",
  year      =  2000,
}

@article{cca-nonlinear5,
  title     = "A neural implementation of canonical correlation analysis",
  author    = "Lai, P L and Fyfe, C",
  journal   = "Neural Networks",
  publisher = "Elsevier BV",
  volume    =  12,
  number    =  10,
  pages     = "1391--1397",
  year      =  1999,
}

@incollection{lid-bailey2021,
  title     = "Relationships between local intrinsic dimensionality and tail entropy",
  author    = "Bailey, James and Houle, Michael E and Ma, Xingjun",
  booktitle = "Lecture Notes in Computer Science",
  publisher = "Springer International Publishing",
  pages     = "186--200",
  series    = "Lecture Notes in Computer Science",
  year      =  2021,
}

@INCOLLECTION{lid-bailey2019,
  title     = "Intrinsic dimension of data representations in deep neural
               networks",
  author    = "Ansuini, Alessio and Laio, Alessandro and Macke, Jakob H and Zoccolan, Davide",
  booktitle = "Proc. NeurIPS",
  publisher = "Curran Associates Inc.",
  year      =  2019
}

@INPROCEEDINGS{lid-Pope2021,
  title   = "The Intrinsic Dimension of Images and Its Impact on Learning",
  author  = "Pope, Phillip E and Zhu, Chen and Abdelkader, Ahmed and Goldblum, Micah and Goldstein, Tom",
  booktitle = "Proc. ICLR",
  year    =  2021
}

@INPROCEEDINGS{lid-Ma2018,
  title    = "Dimensionality-driven learning with noisy labels",
  author   = "Ma, Xingjun and Wang, Yisen and Houle, Michael E and Zhou, Shuo and Erfani, Sarah M and Xia, Shu-Tao and Wijewickrema, Sudanthi and Bailey, James",
  year     =  2018,
  booktitle = "Proc. ICML",
}

@INPROCEEDINGS{lid-Gong2019,
  title     = "On the Intrinsic Dimensionality of Image Representations",
  author    = "Gong, Sixue and Boddeti, Vishnu and Jain, Anil",
  booktitle = "IEEE/CVF Conference on Computer Vision and Pattern Recognition (CVPR)",
  pages     = "3982--3991",
  year      =  2019
}

@inproceedings{zhu2022noise_robust_ssl,
      author       = {Qiu{-}Shi Zhu and
                      Jie Zhang and
                      Zi{-}qiang Zhang and
                      Ming{-}Hui Wu and
                      Xin Fang and
                      Li{-}Rong Dai},
      title        = {A Noise-Robust Self-Supervised Pre-Training Model Based Speech Representation Learning for Automatic Speech Recognition},
      booktitle    = {Proc. {ICASSP}},
      pages        = {3174--3178},
      publisher    = {{IEEE}},
      year         = {2022},
      doi          = {10.1109/ICASSP43922.2022.9747379}
}

@inproceedings{lid-weerasinghe2022,
      author       = {Sandamal Weerasinghe and
                      Tamas Abraham and
                      Tansu Alpcan and
                      Sarah M. Erfani and
                      Christopher Leckie and
                      Benjamin I. P. Rubinstein},
      title        = {Local Intrinsic Dimensionality Signals Adversarial Perturbations},
      booktitle    = {61st {IEEE} Conference on Decision and Control, {CDC} 2022, Cancun,
                      Mexico, December 6-9, 2022},
      pages        = {6118--6125},
      publisher    = {{IEEE}},
      year         = {2022},
      url          = {https://doi.org/10.1109/CDC51059.2022.9992383},
      doi          = {10.1109/CDC51059.2022.9992383},
      bibsource    = {dblp computer science bibliography, https://dblp.org}
}

@inproceedings{lid-Ansuini2019,
    author = {Ansuini, Alessio and Laio, Alessandro and Macke, Jakob H. and Zoccolan, Davide},
    title = {Intrinsic dimension of data representations in deep neural networks},
    year = {2019},
    booktitle = {Proceedings of the 33rd International Conference on Neural Information Processing Systems},
    articleno = {549},
    numpages = {12}
}

@article{lid-amsaleg1028,
  title = {Extreme-Value-Theoretic Estimation of Local Intrinsic Dimensionality},
  author = {Amsaleg, Laurent and Chelly, Oussama and Furon, Teddy and Girard, St{\'e}phane and Houle, Michael E. and Kawarabayashi, Ken-ichi and Nett, Michael},
  year = 2018,
  month = nov,
  journal = {Data Mining and Knowledge Discovery},
  volume = {32},
  number = {6},
  pages = {1768--1805},
  issn = {1573-756X},
  doi = {10.1007/s10618-018-0578-6},
}

@inproceedings{cca-nonlinear6,
  title     = "Unsupervised learning of acoustic features via deep canonical correlation analysis",
  author    = "Wang, Weiran and Arora, Raman and Livescu, Karen and Bilmes, Jeff A",
  booktitle = "Proc. ICASSP",
  publisher = "IEEE",
  pages     = "4590--4594",
  month     =  apr,
  year      =  2015
}

@misc{rankme-garrido-lecun-2023,
      author = {Garrido, Quentin and Balestriero, Randall and Najman, Laurent and LeCun, Yann}, 
      title = {RankMe: assessing the downstream performance of pretrained self-supervised representations by their rank}, year = {2023}, 
      booktitle = {Proceedings of the 40th International Conference on Machine Learning}, 
      articleno = {440}, 
      numpages = {46}, 
      series = {ICML'23}
}

\end{document}